\documentclass[fleqn,usenatbib]{mnras}
\usepackage[english]{babel}
\usepackage[T1]{fontenc}
\usepackage{ae,aecompl}
\usepackage{fancyhdr}
\usepackage{amsfonts}
\usepackage{amsmath}
\usepackage{amssymb}
\usepackage{multicol}
\usepackage{layout}
\usepackage{graphicx}
\usepackage{multirow}
\usepackage{color}
\usepackage{multirow}
\usepackage{times}
\usepackage{natbib}
\def\be{\begin{equation}}
\def\ee{\end{equation}}

\usepackage[utf8]{inputenc}
\usepackage{lmodern}

\definecolor{vale}{rgb}{1.,.5, 0.0}

\newif\ifAMStwofonts
\AMStwofontstrue

\graphicspath{{./fig/}}

\title[Testing MOdified Gravity (MOG) theory and dark matter model in  Milky Way using the local observables ]{Testing MOdified Gravity (MOG) theory and dark matter model in  Milky Way using the local observables}
\author[Davari and Rahvar]{
Zahra Davari$^{1}$ and Sohrab Rahvar$^1$  \\ 
	$^1$ Department of Physics,Sharif University of Technology, P.O.Box 11365-9161, Tehran,
	Iran}
\date{Accepted ?, Received ?; in original form \today}

\pagerange{\pageref{firstpage}--\pageref{lastpage}} \pubyear{2015}

\begin{document}
	\label{firstpage}
	
	\maketitle
	\begin{abstract}	
		In this paper, we have investigated one of the alternative theories to dark matter named MOdified Gravity (MOG) by testing its ability to describe the local dynamics of the Milky Way in vertical and transverse directions with the baryonic matter. MOG is designed to interpret the dynamics of galaxies and cluster of galaxies without the need for dark matter. We use local observational data such as the vertical dispersion, rotation curve, surface density and number density of stars in the Milky Way to obtained the parameters of MOG and the baryonic component of MW by implementing a Bayesian approach to the parameter estimation based on a Markov Chain Monte Carlo method. We compare our results with the dark matter model of MW. The two models of MOG and CDM are able to describe equally well the rotation curve and the vertical dynamics of stars in the local MW. The best values for the free parameters of MOG in this analysis is obtained as $\alpha = 8.99 \pm 0.02 $ and $\mu =0.054\pm 0.005$ kpc$^{-1}$. Also, we obtain the parameters of the generalized gNFW model in the dark matter model. Our best value of bulge mass from MOG is  $(1.06 \pm 0.26)\times10^{10}\rm M_{\odot}$ which is consistent with the estimations form the microlensing observations. \end{abstract}
\section{Introduction} \label{sec:intro}	
The standard model of cosmology $\Lambda CDM$  explains the present experimental data about the accelerated expansion of the universe, the power spectrum of the Cosmic Microwave Background, the formation of the large scale structures and gravitational lensing effects by postulating the existence of the exotic forms of the dark matter (DM) and dark energy (DE)\citep{Scolnic:2017caz,Aghanim:2018eyx,Zhao:2018jxv}. The observational evidence of cosmic acceleration has obtained almost two decades ago. The first observational evidence of dark matter is observed by Zwiky in the cluster of galaxies \citep{Zwicky:1937zza}. Also, the observational evidence for dark matter in the spiral galaxies has been investigated in the 1970s where there was a discrepancy between the observed dynamics and the mass inferred from luminous matter (missing mass problem) \citep{Rubin:1970zza,Rubin:1980zd}. Their observations showed that the orbital velocity of stars around the center of spiral galaxies in the outer regions of galaxies is significantly larger than that of expected due to the Newtonian gravity by the visible matter. This observation provides the necessity for the existence of dark halo structure around the spiral galaxies with the mass of one order of magnitude larger than the visible part of the galaxy. 

Although there are many strong indirect evidences that support the existence of DM, however, no direct and indirect evidence has been found for the existence of dark matter particles \citep{Aprile:2018dbl,Atwood:2009ez}. Despite the difficulty in identifying the dark matter contribution to the total mass density in our Galaxy, stellar kinematics, as a tracer of gravitational potential, is the most reliable observable for gauging different matter components. Moreover, at the scale of galaxies, there is tension between the theoretically expected dark matter distribution and its indirectly observed distribution such as core cusp and satellite problems \citep{Moore:1999gc,Klypin:1999uc,Maleki:2019xya}.

An alternative way to explain the observed dynamics of galaxies and clusters is to adopt a modified theory of gravity. There are numerous proposals for modified gravity as proxies for dark matter in the literature for the regime of galaxies and cosmology, such as MOdified Newtonian Dynamics (MOND)\citep{Milgrom:1983ca,Bekenstein:2004ne}, the Tensor-Vector-Scalar theory (TeVeS) and the Modified gravity (MOG) theory \citep{Moffat:2005si}. In these theories are no additional dark mass component in the structures and dynamics is determined purely by the baryonic matter.
The other observations such as comparing the mass construction from the 
gravitational weak lensing and baryonic distribution of  A1689 cluster, \citet{ Nieuwenhuizen:2016uxv} conclude 
that popular alternative gravity theories like MOND and MOG or TeVeS  cannot fit these two data unless an additional dark matter profile is assumed in the halo of galactic cluster.  Also the ability of Weyl gravity and MOG to inferred acceleration of structures without using the dark matter component is discussed in \cite{Dutta:2018oaj,Islam:2019iua,Negrelli:2018gll}.

In this work, we will take two models of DM and MOG  to study the local dynamics of MW. The main goal of this work is to investigate the radial and vertical dynamics of stars in the local MW and compare them with the observed data. We apply the local Milky Way observables, including the vertical dispersion velocity of stars, the rotation curve, the baryonic surface density, and the stellar disk profile. \\The paper is organized as follows: in the section (\ref{sec1}), we briefly review the DM model and  MOG theory. In section (\ref{sec2}) we explain completely local Milky Way observables. We	perform a statistical analysis in order to constraint the free parameters of the models and compare them to each other in section (\ref{sec3}) and we draw our conclusions in section (\ref{sec4}). 				
\section{Model Definition And Assumptions}\label{sec1}
In this section, we introduce the baryonic component of the Milky Way, the dark halo component and also MOG as a modified gravity to investigate the dynamics of Galaxy using only the baryonic part of the galaxy.  
\subsection{The baryonic component of the Milky Way}
The exact distribution of the (visible) baryonic components in our galaxy is not exactly clear and there are uncertainties \citep{Bland:2017,Lin:2019yux}. We use the mass distribution of the baryonic content (stars and gas) within the Milky Way according to a set of observations inferred the morphologies in three main visible baryonic components \citep{Iocco:2015iia}. 
\begin{itemize}
	\item [1-] {\it Stellar Bulge:} the bulge dominates the inner few kpcs (2-3 kpc) of the  Milky  Way by a  triaxial shape with a bar extending at positive Galactic longitudes and it contains approximately 15\% of the luminous matter. There are some profiles for the bulge as exponential, Gaussian and power-laws. Here we select a Hernquist density profile \citep{Hernquist:1990be}
	\begin{equation}
	\label{1}
	\rho_{\star,b}(r)=\frac{M_{\star,b}}{2\pi}\frac{r_{\star,b}}{r}\frac{1}{(r+r_{\star,b})^3},
	\end{equation}
	where $r_{\star,b}$ is the bulge scale radius and $M_{\star,b}$ is a normalization constant that sets the total mass of the bulge. We fixed the scale radius to $r_{\star,b}=600$ pc \citep{Bland:2017}, while the normalization is allowed to vary. 
	\item[2-] {\it The stellar Disc:} the stellar disk extends up to $\sim$ 20 kpc from the galactic center, presents axial symmetry and contains approximately $\sim$75\% of all Galactic stars. It has been modeled by different authors with the help of surveys of photometric data across the Galaxy \citep{Iocco:2015xga}. In the observational studies, \cite{Misiriotis:2006qq} analyzed COBE dust emission maps to constrain the parameters of the Galactic model. Such a model is comprised of axisymmetric distributions for the stars (composed of a bulge and a disk), the dust (cold and warm) and the gas (molecular $\rm H_2$ and atomic HI). The stellar disk in the cylindrical coordinate is given by double exponential function as follows  
	\begin{equation}\label{rhodisk}
	\rho_{\star,disk}(r,z)=\tilde{\rho}_{\star} \exp(-\frac{r}{h_{\star,R}}-\frac{|z|}{h_{\star,z}}),
	\end{equation}
	where $h_{\star,R}$ and $h_{\star,z}$ are the length scales  of the disk, $\tilde{\rho}_{\star}=\frac{M}{4\pi h_{\star,z}h_{\star,R}^2} $ is the normalization density and M is the mass of disk. We fix the vertical height scale to the best fit value of $h_{\star,z}=300$ pc \citep{Bland:2017}. 
	\item[3-]{\it The interstellar gas:} The third part of baryonic matter in the Milky Way is in the form of gas which can be found in three different forms of molecular, atomic and ionized matter. The interstellar gas is composed mainly by hydrogen (90.8\% by number or 70.4\% by mass), 9.1\% by number (28.1\% by mass) of helium and negligible amounts of heavier elements.  The description of the interstellar gas in the MW is decomposed into two parts: one that describes the interstellar gas within the inner 3 kpc of the Galaxy and another for the distribution of gas beyond 3 kpc from the galactic center. We take a double exponential density profile for the gaseous disk, also alike the equation (\ref{rhodisk}). We fixed the scale height of the disk of gaseous component to $h_{g,z}=130$ pc and supposed $h_{g,R}=2h_{\star,R}$ \citep{Bovy,Binney2008}. 
\end{itemize}
\subsection{Dark matter model for halo} \label{sec1-1} 
The nature of the dark matter (DM) particle is unknown. We can infer some of its properties from its gravitational effects to explain the growth of structures with non-relativistic particles so-called cold dark matter. From the dynamics of the spiral arm and nearby galaxies such as large and small Magellanic clouds, we can infer that halo is extended up to  $\approx 200$ kpc. Furthermore, the mass of the halo of DM is one order of magnitude larger than the mass of stars and gas from the dynamics of MW \citep{Benito:2019vef}.
There are various models have been proposed for the distribution of DM Halo density profiles for our galaxy such as the isothermal, the Navarro-Frenk-White (NFW) \citep{Navarro:1995iw} and Einasto profiles \citep{Einasto:1965czb}. The generalized Navarro-Frenk-White (gNFW) profile is a generalization of the NFW profile \citep{Navarro:2003ew}  and a good description of the mass distribution of DM halos from DM-only simulations from dwarf galaxies up to galaxy clusters which takes into account the unknown inner density profile of DM halos and it represents a cuspy profile diverges towards smaller $r$ values. This function is given by 
\begin{equation}\label{eqnfw}
\rho_{DM}(r)=\frac{\tilde{\rho}_{dm}}{(\frac{r}{r_s})^{\gamma}(1+\frac{r}{r_s})^{3-\gamma}}=\tilde{\rho}_0(\frac{R_{0}}{r})^{\gamma}(\frac{r_s + R_0 }{r_s +r })^{3-\gamma},
\end{equation}
where  $\gamma$ is the inner slope of the density profile, $r_s$ is the scale radius, $\tilde{\rho}_{dm}$ is the characteristic density and $\tilde{\rho}_0$ is the local density of halo.
In work we adopt gNFW profile and $\tilde{\rho}_{dm}, \gamma$ and $r_s$ are considered as the free parameters. Typical ranges from the N-body simulations are $0.9\ll \gamma \ll 1.2$ and $r_s=19\rm$ kpc \citep{Lisanti:2018qam}.
\subsection{MOG Model} \label{sec1-2}
We will use the MOdified Gravity theory (MOG) also called Scalar-Tensor-Vector-Gravity  \citep{Moffat:2005si} to examine the local dynamics of Galaxy and compare the results with that of dark matter theory.  MOG is a relativistic modified gravity theory with a tensor field to represent the gravity field and the extra degrees of freedom with a massive vector field $\phi^{\mu}$ and the three scalar fields of G, $\mu, \omega$ which represent the gravitational coupling strength, the mass of the vector field and its coupling strength, respectively.
The action of this theory is given by $S = S_G  +S_\phi + S_S + S_{matter}$ where $S_{matter}$ is the action of matter field and 
\begin{eqnarray}
&&S_G=-\frac{1}{16\pi}\int \frac{1}{G}(R+2\Lambda)\sqrt{-g}d^4x,\\
&&S_{\phi}=-\frac{1}{4\pi}\omega \int [\frac{1}{4}B^{\mu\nu}B_{\mu\nu}-\frac{1}{2}\mu^2\phi_{\mu}\phi^{\mu}\\\nonumber
&&\qquad+V_{\phi}(\phi_{\mu}\phi^{\mu})]\sqrt{-g}d^4x,\\
&&S_S=-\int\frac{1}{G}[\frac{1}{2}g^{\alpha \beta}(\frac{\nabla_{\alpha}G\nabla_{\beta}G}{G^2}+\frac{\nabla_{\alpha}\mu\nabla_{\beta}\mu}{\mu^2})+\\\nonumber
&&\qquad\frac{V_G(G)}{G^2}+\frac{V_{\mu}(\mu)}{\mu^2}]\sqrt{-g}d^4x,
\end{eqnarray}
where $B_{\mu\nu}=\partial_{\mu}\phi_{\nu}-\partial_{\nu}\phi_{\mu}$ is the Faraday tensor of the vector field and $\nabla_{\nu}$ is the covariant derivative with respect to the	 metric $g_{\mu\nu}$ and $V_{\phi} (\phi_{\mu} \phi^{\mu} )$, $V_G(G)$ and $V_{\mu} (\mu)$ are the self-interaction potentials associated with the vector field and the scalar fields, respectively.
On astrophysical scales for studying the behavior of MOG,  we can use the weak field approximation for the dynamics of gravitating systems with perturbing them around Minkowski space for the arbitrary distribution of non-relativistic matter. Under this assumption, the scalar fields remain constant and the acceleration of a test particle can be written as,
\begin{equation}
\label{mog}
\vec{a}(\vec {x})=-G_N\int\frac{\rho(\vec {x'})(\vec {x}-\vec {x'})}{|\vec {x}-\vec {x'}|^3}\times[ 1+\alpha-\alpha e^{-\mu|\vec {r}|}(1+\mu|\vec {r}|)]d^3\vec {x'}.
\end{equation}
The parameter $\alpha$ and the vector field mass $\mu$ control the strength and the range of the modified gravity. We note that for a point mass object with the mass of $M$, these two parameters are given as \citep{Moffat:2007nj}: 
	
\begin{equation}\label{equE}
\alpha=\frac{M}{(\sqrt{M}+E)^2}\left(\frac{G_{\infty}}{G_N}-1\right),
\end{equation}	
 	and	
\begin{equation}\label{equD}
\mu=\frac{D}{\sqrt{M}},
\end{equation}	
where $G_{\infty}\sim 20G$ represents the effective gravitational constant at infinity, while D and E are determined using observational data. Since we are studying the extended objects in the weak approximation of this theory, the $\alpha$ and $\mu$ parameters depend on the mass and the coupling 
constant of the vector field with the matter field and they are considered as constant 
parameters \citep{rahvar1}.
 The numerical values of these parameters have been reported by fitting the rotation curves of galaxies to observational data as well as the 
X-ray emission from the cluster of galaxies to be  $\alpha = 8.89$ and $\mu= 0.042$ $\rm kpc ^{-1}$ \citep{rahvar1,rahvar2}.  
\section{Data Sets}\label{sec2}
In the section, we explain the observational parameter that we will use to investigate the CDM and MOG. Here, we use the Galactocentric cylindrial system $(R, \phi, z)$ where $R$ is the projected Galactocentric distance, $\phi$ in the direction of Galactic rotation and z towards the North Galactic Pole. One of the key parameters to interpret any other kinematic observations of the Milky Way
is the distance from the Sun to the Galactic Center, $R_0$. There is still
significant uncertainty on this parameter. In this work, we adapt the solar position at $R_0=8.12$ kpc and at the Galactic mid-plane ($z\simeq 0$ kpc)
as the fiducial values, consistent with the observation of Sgr A$^\star$
black hole \citep{Abuter:2018drb,Eilers_2019}.	
\subsection{Rotational Curve of Galaxy} 
Rotation Curve  is one of the main evidences for the problem of missing mass
in the sprial galaxies. For rotatonal velocity well beyond the galaxy core, we can obtain the acceleration of a test particle from the gradient of the potential, $a=-\nabla \Phi_{eff}$ as follows
\begin{equation}
\vec{a}(\vec {x})=-G_N\int\frac{\rho(\vec {x'})(\vec {x}-\vec {x'})}{|\vec {x}-\vec {x'}|^3}\times f(\vec {x}-\vec {x'})d^3\vec {x'},
\end{equation}
where for CDM, $\rho(r)=\rho_B(r)+\rho_{DM}(\vec {r})$ and $f = 1$ and for MOG,  $\rho(r)=\rho_B(r)$ and  $f(\vec{r})=1+\alpha-\alpha e^{-\mu|\vec {r}|}(1+\mu|\vec {r}|)$. From the density profile, we can infer the corresponding gravitational potential and therefore, circular velocities. This is the algorithm we use for 
calculating the rotation curve of Galaxy:  
\begin{equation}
\rho_i(\vec{x})\rightarrow\Phi_i(\vec{x})\rightarrow v_i(R);
\end{equation}
where $\vec{x}$ is a three-dimensional vector and R is the distance to the center of Galaxy.
In the weak gravitational limit, the gravitational potential relates to the density profile with the boundary condition $\Phi_i\rightarrow 0$ to $|x|\rightarrow \infty$ where the rotation velocity obtain as 
\begin{equation}
	v_i^2(r,z)=r G \int\int\int\frac{\rho_i(r',z')f(|\vec {r}-\vec {r'}|)}{|\vec {r}-\vec {r'}|^3}(r-r'\cos\phi')r'dr'd\phi'dz',
\end{equation}
where $|\vec {r}-\vec {r'}|=\left(r^2+r'^2-2rr'\cos\phi'+(z-z')^2\right)^{1/2}$. Measurement of the RC of the Milky Way is much harder than external galaxies. This is due to our interior position that complicates some measurements, such as the extended RC of the gas in the disk.\cite{Eilers_2019} determined the circular velocity at the Sun's Galactocentric radius with its formal uncertainty to be $v_c (R_0 )= 229.0 \pm 12 $ km s$^{-1}$ . They found that the velocity curve is gently but significantly declining at $-1.7 \pm 0.47$ kms$^{-1}$ kpc$^{-1}$, with a systematic uncertainty of $0.46$ km s$^{-1}$ kpc$^{-1}$, beyond the inner 5 kpc. The value of the circular velocity at the Sun's Galactocentric radius has an important role to constraint the mass distribution of our Galaxy and the local dark matter density and we will use these values in our main and second analysis. The RC can also be used to construct the realistic Galactic mass model by fitting the RC with a parameterized multi-component Milky Way, consisting of, for instance, a bulge, a disc, and a dark matter halo.
\subsection{Number Density and Vertical Velocity Dispersion}
In this section, we use the number density of stars in the Galactic disk. \cite{Kuijken:1989hu} introduced a technique to determine the integral surface mass density of the disc near the Sun. They had used K-dwarfs as the ideal stars in SDSS/SEGUE catalog where they stated that K-dwarfs in the present context are in dynamical equilibrium in Galactic potential as phase-mixed and their distance can be well determined. These stars also can be found to $|z|>1$ kpc, therefore, we make sure to measure the total surface mass density. In this direction, \cite{Zhang:2012rsb}  derived the spatial and velocity distribution by using a sample of 9000 K-dwarf. The distribution of the K-dwarf sample is categorized into subsets that are abundance-selected in the $[\alpha/Fe]$ vs. $[Fe/H]$ plane:
 \begin{itemize}
 	\item metal-rich: $[Fe/H]\in[-0.5, 0.3]$, $[\alpha/Fe] \in [0., 0.15]$.
 	\item intermediate metallicity: $[Fe/H] \in [-1.0, -0.3]$, $[\alpha/Fe] \in [0.15, 0.25]$.
 	\item metal-poor: $[Fe/H] \in [-1.5, -0.5]$, $[\alpha/Fe] \in [0.25, 0.50]$.
 \end{itemize}
 These sub-samples contain 3672, 1416 and 2001 stars, respectively. In order to constrain the gravitational vertical force in the vicinity of the Sun by measuring the mean number density of stars and observed vertical velocity dispersion profiles, we can use the Jeans equations. Hence we model the number density profile of each sub-population $n_i(z)$ as a simple exponential of unknown scale height $h_i$ as
 \begin{equation}
 n_i(z)=\tilde{n_i}\exp(-|z|/h_i),
 \end{equation}
 where index "i" pointed to three tracer populations. From the vertical z-Jeans equation in the cylindrical coordinate and  for a  steady-state disk, 
 \begin{equation}
 \label{jeans}
 \frac{1}{R}\frac{\partial}{\partial R}[Rn_i(z) \sigma ^2_{Rz}]+\frac{\partial}{\partial z}[n_i(z)
  \sigma ^2_{zz,\rm i}(z)]+n_i(z) \frac{\partial \Phi_{\rm eff}}{\partial z}=0,
 \end{equation} 
where $\sigma_{\rm ij}=< v_i v_j>$ is the velocity dispersion tensor and $ \sigma ^2_{zz,\rm i}$ and $ \sigma ^2_{\rm Rz}$ are the velocity dispersion components of a tracer population moving in the vertical gravitational potential $\Phi _{\rm eff}$. 
The first term in equation (\ref{jeans}) is called the 'tilt' term \citep{Zhang:2012rsb} and for an almost symmetric disk with respect to the Galactic plane, we can ignore the first term of the Jeans equation. 
Therefore the dominant gradient in the Jean's equation is the z-gradient of the disk  and the vertical Jeans equation simplifies as 
\begin{equation}\label{eqnz}
\frac{\partial}{\partial z}[n_i(z)\sigma_{zz,\rm i}(z)^2]=-n_i(z)\frac{\partial\Phi_{\rm eff}(z,R)}{\partial z}.
\end{equation}
We solve this equation in the Sun’s vicinity ($R = R_0$).
In addition to  the Boltzmann equation, we use the Poisson equation in cylindrical coordinate to connect  theory to the observations as it relates the potential to the density in terms of $z$ as:
\begin{equation}
\frac{\partial ^2 \Phi_{\rm eff}}{\partial z^2}=4 \pi G[\rho_B(z)+\rho_{DM}^{\rm eff}(z,R_0)],
\end{equation}
where the effective dark matter density \citep{Garbari:2012ff} by moving the radial component of the Poisson equation to the right-hand side is 
\begin{equation}
\rho_{DM}^{\rm eff}(z,R_0)=\rho_{DM}(z,R_0)-\frac{1}{4\pi Gr}\frac{\partial v^2_c(R_{0})}{\partial r},
\end{equation}
and the second term at the right-hand side of the equation associate to the rotation curve at the position of the Sun. This term can be calculated from the Oort constants \citep{Binney1998} of $A$ and $B$  as
\begin{equation}
\frac{1}{4\pi Gr}\frac{\partial v^2_c(r)}{\partial r}=\frac{B^2- A^2}{2\pi G},
\end{equation}
where $A  = \frac{1}{2}(\frac{v_c}{r}-\frac{dv_c}{dr}){(R_0)}$ and $B =- \frac{1}{2}(\frac{v_c}{r}+\frac{dv_c}{dr}){(R_0)}$. 
There is a wide variety of measurements of these constants, but in this study we used  one of the  most accurate and recent values of these measurements from \cite{Bovy} and \cite{Schutz:2017tfp} as $ A =15.3 \pm 0.4 \rm km s^{-1}$ $ \rm kpc^{-1}$  and $B =-11.9 \pm 0.4$ $\rm km s^{-1} \rm kpc^{-1}$ .\\
We rewrite equation (\ref{eqnz}) in the integral form as 
\begin{equation}\label{eqsz}
\sigma_{zz,i}(z)^2=\frac{-1}{n_i(z)}\int_z^\infty n_i(z')\frac{\partial  \Phi_{\rm eff}}{\partial z'}	dz'.
\end{equation}
The upper bound of this integral goes up to infinity, however, the integrand falls off exponentially in the z-direction and from our numerical calculation, we put cut-off a the altitude of  $z = 2$ kpc.	
The dispersion velocity of stars, as well as the number density of the stars, are the observable parameters. In Figure (\ref{fig-5}), we show the observational data of  $\sigma_{zz}(z)$ and $ n(z) $ in terms of $|z|$ for each of the three tracer populations.\\ 
It is well known that the velocity dispersion of a stellar population $\sigma_{zz}(z)$ increases with age or deficit of the metallicity \citep{Wielen:1977zz,Rebassa_Mansergas_2016} as seen in Figure (\ref{fig-5}). Moreover, the dispersion velocity of metal-poor stars is almost twice of the intermediate or metal-rich stars.  
The vertical velocity dispersion increases only for the metal-poor stars with distance from the plane. The other types of stars have the non-monotonic profile of the dispersion velocity as a function of $z$. 
\begin{table*}
	\centering
	\caption{Marginalized constraints (1$\sigma$) uncertainties of free parameters using the Gaia DR2 data.}
	\begin{tabular}{|c | c| c|  c|c |c |c|  c|c |c |c |}
		\hline \hline
		$\rm Parameter$	&$\tilde{\rho}_*$& $\tilde{\rho}_g$& $h_{*,R} $& $M_{*,b}$& $\tilde{\rho}_{dm}$& $\gamma$&$R_s$&$\alpha$&$\mu$&$\chi^2$\\
		\hline
		$\rm Unit$	&$M_{\odot}\rm pc^{-3}$& $M_{\odot}\rm pc^{-3}$& $\rm kpc $& $10^{10}M_{\odot}$& $M_{\odot}\rm pc^{-3}$& $-$&$\rm kpc$&$-$&$\rm kpc^{-1}$&$-$\\
		\hline							
		$\rm DM$&$ 1.08\pm0.40$& $0.46\pm0.31$&$ 2.98\pm0.52$& $0.23\pm0.30$& $0.03\pm 0.02$& $0.63\pm0.28$& $37.5\pm4.50$& -&-&$4.07$\\
		\hline
		$\rm MOG$&$1.99\pm0.49$& $0.20\pm0.17$&$ 2.81\pm0.29$&$ 0.36\pm0.28$&$-$&$-$&$-$&$ 8.03\pm0.05$&$ 0.05\pm0.01$&$4.10$\\
		\hline \hline
	\end{tabular}\label{tabvr38}
\end{table*}
\subsection{Density profile of baryonic matter}
In this study, we need the numerical value of the baryonic matter density of galaxy in the Galactic bulge and disk. It is more convenient to use the surface density rather than the volume density of matter. The surface density of baryonic matter can be defined as 
\begin{equation}\label{siggas}
			\Sigma_{j}(R_{0})=2\int_{0}^{z_{max}}\rho_j(R_{0},z')dz',
\end{equation}
where $j$ represents the two components of the baryonic matter as the gas and stars and the column density is measured around the Sun. 
We use the measured values of $\Sigma_{g,obc}=12.6\pm1.6~M_{0}\rm pc^{-2}$ and $\Sigma_{\star,obs}=31.2\pm1.6~M_{0}\rm pc^{-2}$ where the maximum distance of $z_{max} = 1.1$ kpc is taken for the integration \citep{McKee:2015hwa}. 
These data obtained from direct photometric observations \citep{Lisanti:2018qam}.
\section{Set up and Methodology}\label{sec3}
\begin{figure}
	\begin{center}
		\includegraphics[width=8cm ]{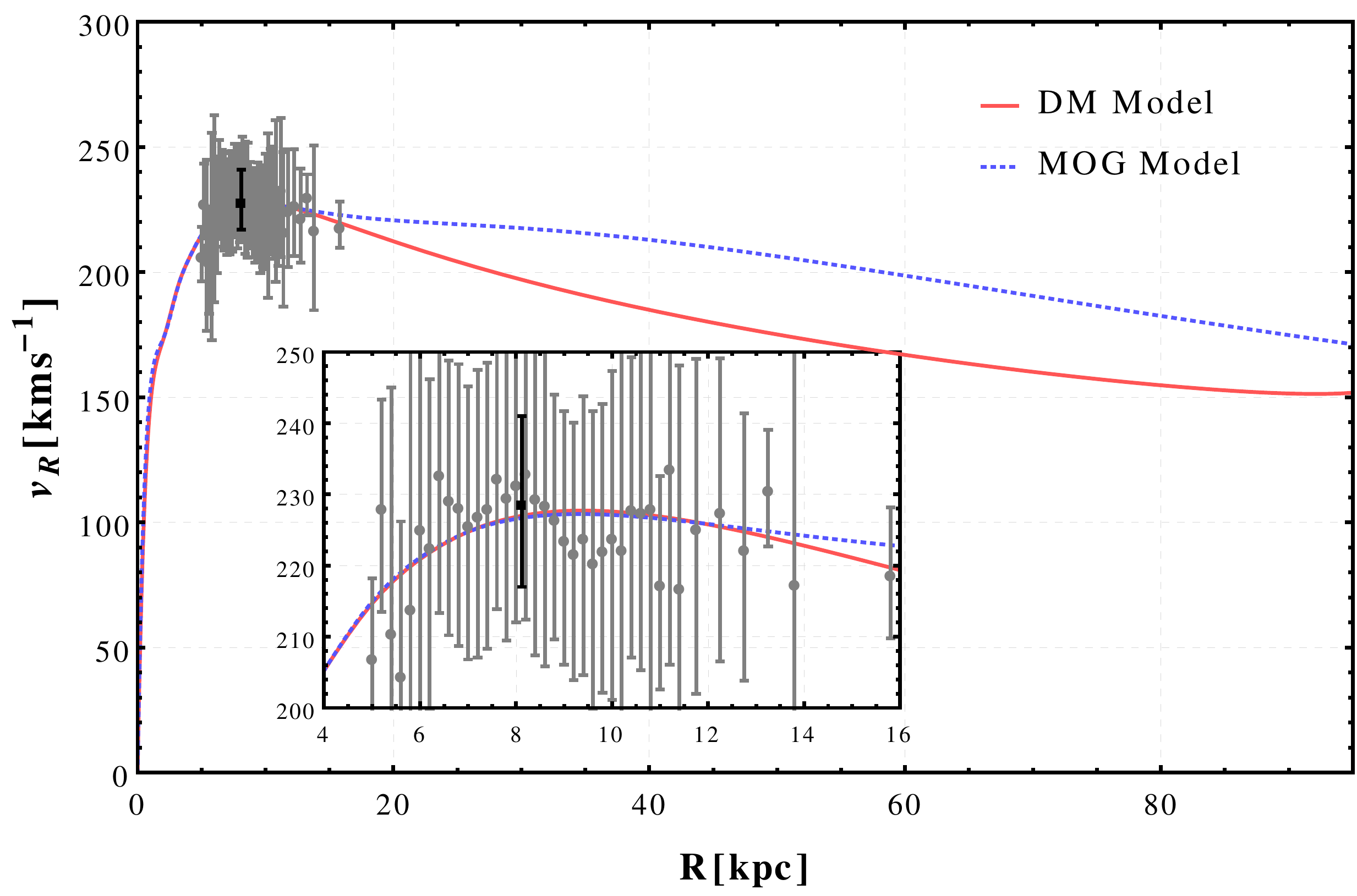}
		\caption{Rotation curves for DM (solid line) and MOG (dashed line) models according to values given in the Table (\ref{tabvr38}). Gaia DR2 data have been shown with their error bars.}
		\label{fig-1}
	\end{center}
\end{figure}

In this section, we want to constrain the free parameters of the	DM and MOG models using the observational data via the Bayesian likelihood analysis. The main purpose of this study is to compare these models with local Milky Way observables. For the first step, we use "Rotation Curve data" around the location Sun (i.e $R_0 = 8.12$ kpc) to compare the two models.\\
This method may give us an overall view at the beginning of this study.
We use 39 data points across Galactocentric distances of $5<R<16$
kpc from the second Gaia data release (DR2) which is shown in Figure (\ref{fig-1}), according to Table (3) in \cite{Crosta:2018var}. Gaia DR2 has provided extremely well measured proper motions and positions on the sky for a significantly large volume of our Galaxy, as well as precise parallaxes within $\sim$2-3 kpc from the solar neighborhood. We can write the likelihood function for the rotation curve as
\begin{equation}
L ({\bf p})\propto \exp\left[-\frac{1}{2}\sum_{i}^N\left(\frac{v_R(R_i,{\bf p})-v_{R,i}}{\sigma_{v_{R,i}}}\right)^2\right], \,
\end{equation}
where the free vector ${\bf p}$ is $\{\tilde{\rho}_*,\tilde{\rho}_g,h_{*,R},M_{*,b},\tilde{\rho}_{dm},\gamma,R_s\}$ and $\{\tilde{\rho}_*,\tilde{\rho}_g,h_{*,R},M_{*,b},\alpha,\mu\}$ for DM and MOG models, respectively.
We consider uniform priors on these parameters and the range allowed for each parameter is as follows 
\begin{eqnarray}
&&\tilde{\rho}_* \in [0,3]M_\odot pc^{-3},\tilde{\rho}_g \in [0,0.5]M_\odot pc^{-3},\nonumber\\
&& h_{*,R} \in [0.1,4]\rm kpc, M_{*,b}\in [0.1,4]10^{10}M_\odot,  \nonumber\\
&&\tilde{\rho}_{dm} \in [0.01,0.08]M_\odot pc^{-3},\gamma \in [0,2],R_s \in [5,100]\rm kpc, \nonumber\\
&&\alpha \in [7,16], \mu \in[0.03,0.07]\rm kpc^{-1} \nonumber
\end{eqnarray}
and by using the Markov Chain Monte Carlo (MCMC) implementation we recover the posterior distributions for the parameters of both models. We summarize the numerical results of this likelihood analysis in Table (\ref{tabvr38}). To quantitatively asses these models, we compare the best fitting to these two models using the reduced $\chi^2$, the Akaike Information  Criterion (AIC) and the Bayesian information criterion (BIC) formulas as given by:
\begin{align}
	\chi^2_{\rm red} &= \frac{\chi^2_{\rm min}}{N-M} \,, \\
	{\rm AIC}&=\chi^2_{\rm min}+2M+\frac{2M(M+1)}{N-M-1} \,, \label{aic1}\\
	{\rm BIC}&=\chi^2_{\rm min}+M\ln N \,, \label{bic}
\end{align}
where $N$  is the number of data points and $M$ is the number of free parameters in each model. The preferred model amongst the other models corresponds to the minimum AIC and BIC values. For comparing the other models with the best model, we use two parameters of $\Delta \rm AIC$ and $\Delta \rm BIC$. These parameters are interpreted according to the calibrated Jeffreys' scales shown in Tables 2 and 3 in \cite{Rivera:2016zzr}.
In comparison between the dark matter model and MOG with the rotation curve of the Milky Way galaxy, we obtain the following results of 
\begin{itemize}
		\item [$\bullet$] For DM model: ${\chi}^2_{\rm red}=0.127$, $N=7$, $\rm AIC=21.7$ and $\rm BIC=29.7$.
		\item [$\bullet$] For MOG model: ${\chi}^2_{\rm red}=0.124$, $N=6$, $\rm AIC=18.8$ and $\rm BIC=26.1$.
\end{itemize}
\begin{table*}
	\centering
	\caption{Marginalized constraints (1$\sigma$) uncertainties of free parameters using dynamical local observational.}\label{tabbest}
	\begin{tabular}{|l|l|l|l|}
		\hline\hline
	$Parameter$	&$\rm Unit$	&$\rm DM$&$\rm MOG$\\
		\hline
		$\tilde{\rho}_*$&$M_{\odot}\rm pc^{-3}$&$ 0.57^{+0.12}_{-0.12}$&$2.21^{+0.30}_{-0.27}$\\
		\hline
		$\tilde{\rho}_g$&$M_{\odot}\rm pc^{-3}$&$0.16^{+0.10}_{-0.10}$&$0.31^{+0.11}_{-0.11}$\\
		\hline
		$h_{*,R} $&$\rm kpc $&$ 3.43^{+0.12}_{-0.12}$&$2.20^{+0.14}_{-0.14}$\\
		\hline
		$M_{*,b}$& $10^{10}M_{\odot}$&$0.61^{+0.11}_{-0.11}$&$ 1.06^{+0.21}_{-0.26}$\\
		\hline
		$\tilde{\rho}_{dm}$& $10^{-2}M_{\odot}\rm pc^{-3}$&$1.37^{+0.16}_{-0.16}$&$-$\\
		\hline
		$\gamma$&$-$&$1.00^{+0.02}_{-0.02}$&$-$\\
		\hline
		$R_s$&$\rm kpc$& $50.93^{+0.56}_{-0.56}$&$-$\\
		\hline
		$\alpha$&$-$&$-$&$ 8.99^{+0.02}_{-0.03}$\\
		\hline
		$\mu$&$\rm kpc^{-1}$&$-$&$ 0.05^{+0.01}_{-0.01}$\\
		\hline
		$n_i$&$10^{-3}\rm pc^{-3}$&\begin{tabular}[c]{@{}l@{}}$2.15^{+0.05}_{-0.05}$\\ $30.49^{+0.11}_{-0.11}$\\ $4.47^{+0.02}_{-0.02}$\end{tabular}&\begin{tabular}[c]{@{}l@{}}$2.11^{+0.03}_{-0.03}$\\ $28.49^{+0.14}_{-0.17}$\\ $4.29^{+0.02}_{-0.02}$\end{tabular}\\
		\hline
		$h_i$&$\rm kpc$&\begin{tabular}[c]{@{}l@{}}$0.92^{+0.02}_{-0.02}$\\ $0.28^{+0.01}_{-0.01}$\\ $0.47^{+0.01}_{-0.01}$\end{tabular}&\begin{tabular}[c]{@{}l@{}}$0.83^{+0.02}_{-0.02}$\\ $0.28^{+0.02}_{-0.02}$\\ $0.47^{+0.02}_{-0.02}$\end{tabular}\\
		\hline\hline
	\end{tabular}
\end{table*}	

\begin{figure*}
	\begin{center}
		\includegraphics[width=18cm ,height=18cm]{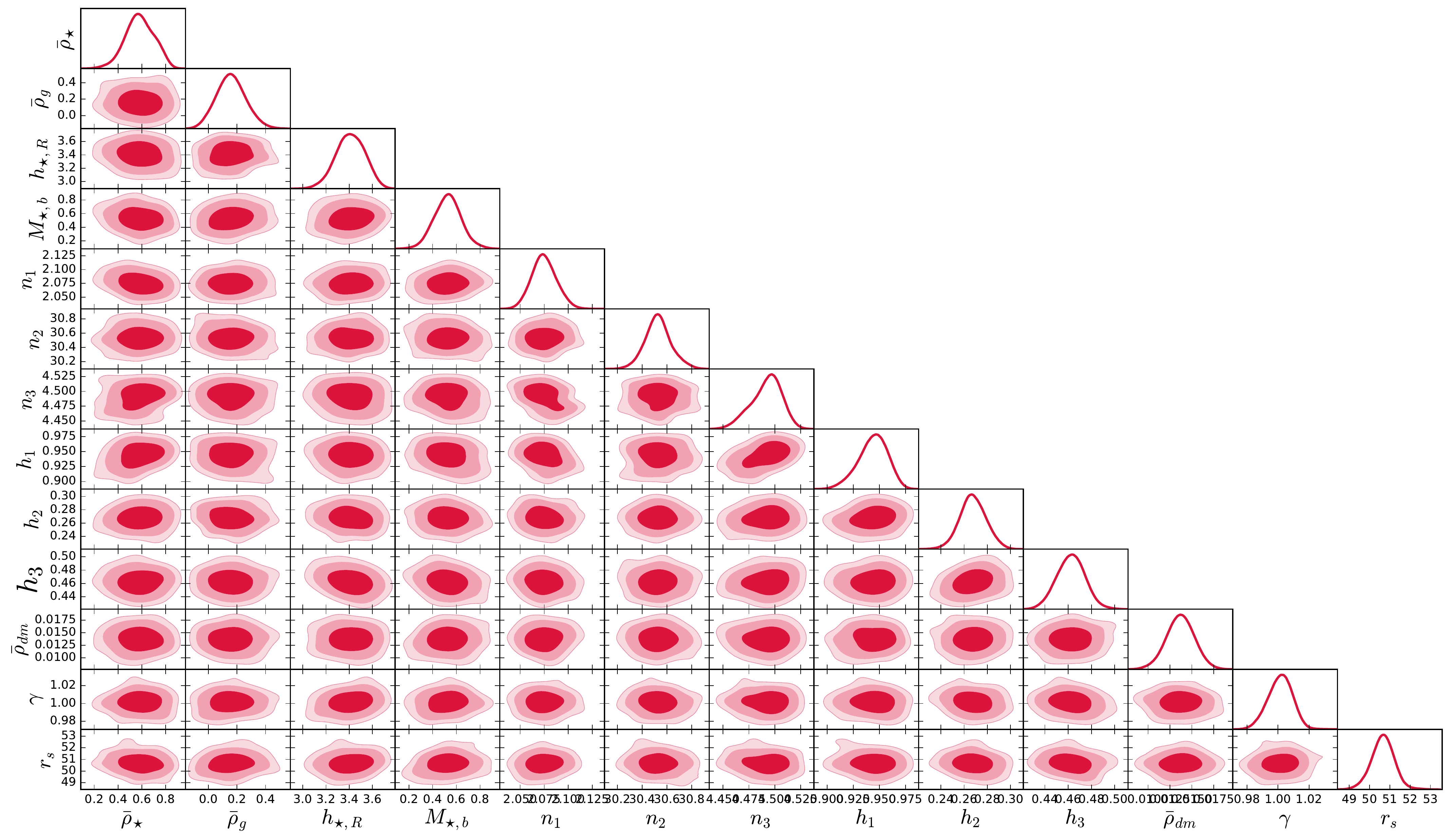}
		\caption{Marginalized 1-, 2- and 3$\sigma$ constraints on the DM model. See table (\ref{tabbest}) for the	numerical values.}
		\label{condmtot}
	\end{center}
\end{figure*}

\begin{figure*}
	\begin{center}
		\includegraphics[width=18cm ,height=18cm]{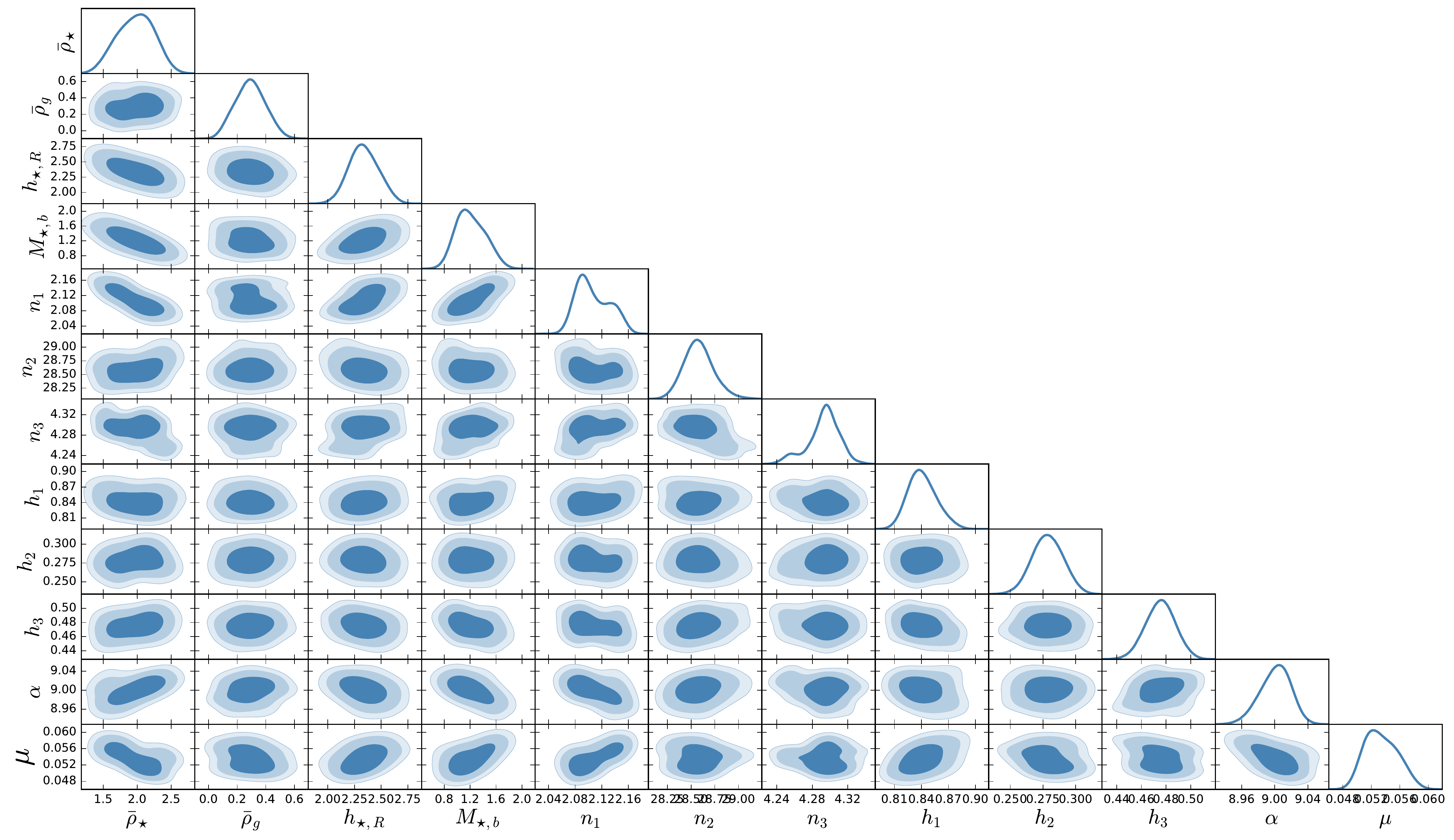}			
		\caption{Marginalized 1-, 2- and 3$\sigma$ constraints on the MOG model. See Table (\ref{tabbest}) for the	numerical values.}
		\label{conmogtot}
	\end{center}
\end{figure*}

By qualitative interpretation of $\Delta \rm AIC$ and $\Delta \rm BIC$, the above results show that these two models are consistent with the observed data.
In Figure (\ref{fig-1}), we compare the RCs of two models with observational Gaia DR2 data by using the best-fitting parameters. The result of the two models is consistent with the observed data points and also, two models show a declining trend by increases in the distance from the center of the galaxy as observed in pervious literatures \citep{10.1093/pasj/64.4.75,10.1093/mnras/stw2096}.\\	
One of the parameters we obtained from the fitting is the mass of the bulge in Table  (\ref{tabvr38}). There are other observational method such as photometric and microlensing observations where the mass of bulge 
has been obtained as $M_{*,b,obs} = (0.62 \pm 0.31)\times 10^{10} \rm M_{\odot}$ from photometric observations \citep{LopezCorredoira:2006na} and  $M_{*,b,obs} = (1.50 \pm 0.38)\times 10^{10} \rm M_{\odot}$ from microlensing observations \citep{Novatin:2007dd}. Our results are consistent with the photometric observations. Also, we obtained the scale length of the stellar disk for two models which are consistent with the observed value of  $h_ {*,R,obs} = 2.6 \pm 0.5$kpc in \cite{Bland:2017}. We note that \cite{Negrelli:2018gll} have performed a test of MOG theories within MW by using the compilations of halo star data from \cite{10.1093/mnras/stw2096} and  tracers {\it galkin}\footnote{https://github.com/galkintool/galkin} released in \cite{Pato:2017yai} for the observed Rotation Curve  up to $200$ kpc by considering the set of different morphologies. They concluded that in none of its present formulation, the MOG theory is able to explain the observed Rotation Curve of the Milky Way.\\		
In following we perform the main constrain comparing the models with the complete set of local Milky Way observables as (i) rotation curve of the disk, (ii) surface density of baryonic matter and (iii) dispersion velocity of local stars.  In another word, the observables are 
\begin{equation}\label{eqchi}
X_{obs}=[(v_{\rm R})_{R_0},(dv_R/dR)_{R_0},\Sigma _{*}^{1.1},\Sigma_{g}^{1.1},n_{\rm i}(z),\sigma_{\rm zz, i}(z)].
\end{equation} 
Here we fix the three galactic parameters of $r_ {*,b}, h_{*,z} $ and $h_{g,z}$ in equation (\ref{rhodisk}) from the observation in \cite{Bland:2017}, \cite{Bovy}. We let the the baryon potential parameters $\{\tilde{\rho}_*,\tilde{\rho}_g,h_{*,R},M_{*,b}\}$ and SEGUE tracer population parameters $\{n_i,h_i | i = 1, 2, 3\}$ to be free parameters. Here, "i" presents three tracer populations consisting of metal-rich, metal-intermediate and metal-poor stars, respectively. Also we let the DM halo model of $\{\tilde{\rho}_{dm},\gamma,R_s\}$ and  for the	MOG case $\{\alpha,\mu\}$ to be free parameters.\\
\begin{table}
	\centering
	\caption{The result of Bayesian Analysis based on a MCMC method.}					
	\begin{tabular}{|c | c| c|  c|c |c|}
		\hline \hline
		$\rm Model$&$\rm N$	&$\chi^2_{\rm min}$& $\chi^2_{\rm red}$& $AIC$& $BIC$\\
		\hline
		$\rm DM$&$ 13$& $39.3$&$ 1.36$& $78.3$& $87.89$\\
		\hline
		$\rm MOG$&$12$& $43.02$&$ 1.43$&$77.8$&$87.87$\\
		\hline \hline
	\end{tabular}\label{tabaictot}
\end{table}
\begin{figure}
	\begin{center}
		\includegraphics[width=8cm ]{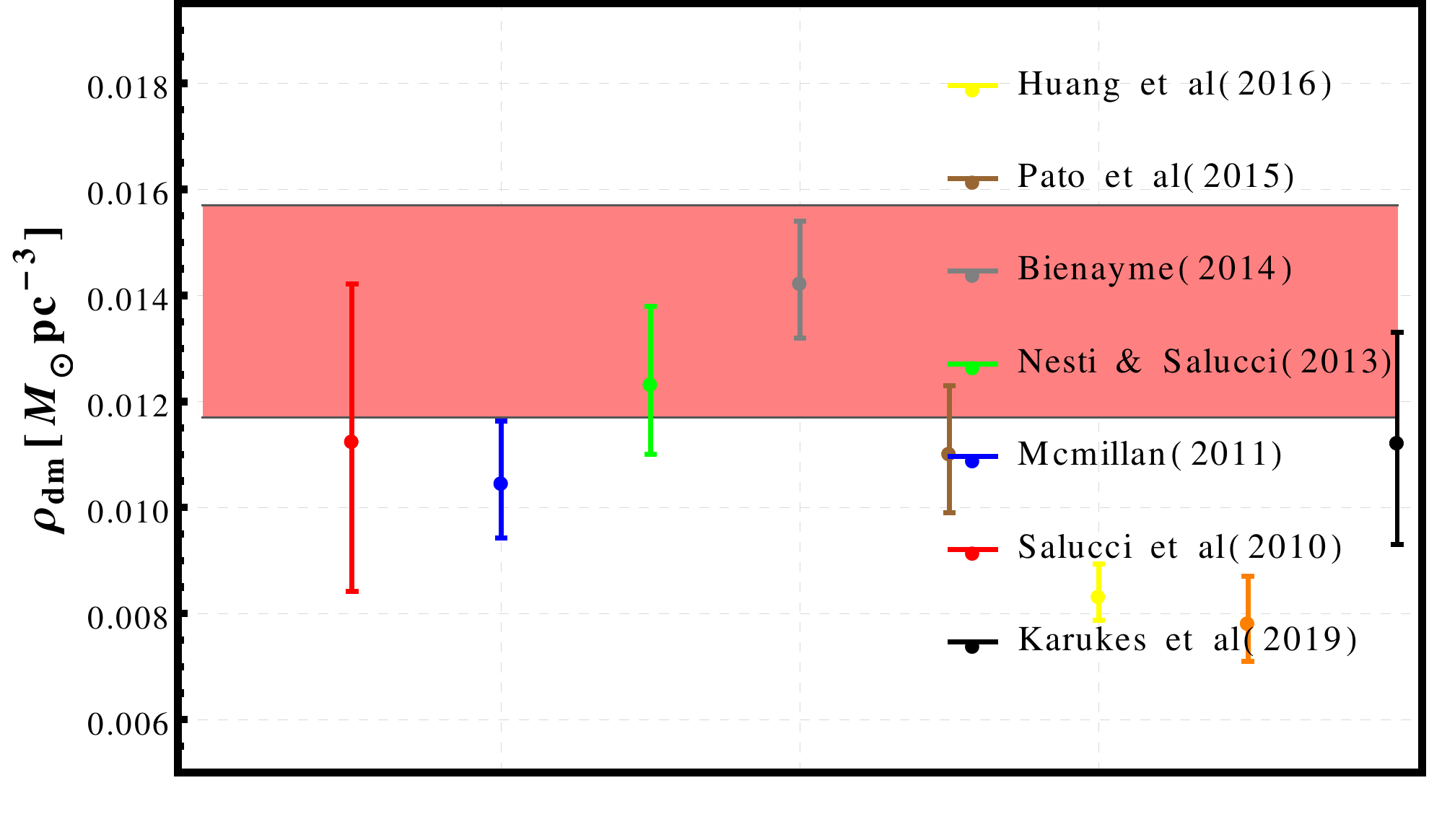}
		\caption{Value of $\tilde{\rho}_{\rm dm}$ found in this work (coloured band) compared with the values obtained in \citep{Salucci:2010qr}, 						\citep{McMillan:2011wd}, \citep{Nesti:2013uwa},\citep{Bienayme:2014kva}, \citep{Pato:2015dua}, \citep{10.1093/mnras/stw2096} ,  \citep{Eilers_2019} and \citep{Karukes:2019jxv}.Note that $1 \rm GeV cm^{-3}= 0.027\rm M pc^{-3}$.}
		\label{fig-2}
	\end{center}
\end{figure}  
We implement the Bayesian approach to the parameter estimation based on an MCMC method.
In Table (\ref{tabbest}) we report the best fit estimate as the mean of the posteriors and their $68\%$ confidence interval. Figures (\ref{condmtot}) and (\ref{conmogtot}) present the $1-3\sigma$ combined likelihood contours for the two models. In Table (\ref{tabaictot}), we report $\chi^2_{\rm red}$ for DM and MOG from this analysis. 
Since we obtained $\Delta \rm AIC=0.5$ and $\Delta \rm BIC=0.02$ so we found that both DM model and MOG theory are consistent with the combination of the rotational velocity and vertical motion of nearby stars in the Milky Way.This compatibility is also shown in \cite{Moffat:2014pia}, except that they assumed MW to be a point particle and used only the rotation velocity data upto a distance of 180 kpc.

We also note that we let the local density of dark matter as the free parameter in gNFW model and obtain it from the best fit to the kinematics data of stars. This data is the combination of the rotation curve of the galaxy and vertical kinematics of a selected group of tracer stars.
We obtain the best value of the dark mater density in equation (\ref{eqnfw}),  $\bar{\rho}_{dm}=0.0137\pm 0.002 \rm ~M_{\odot}\rm pc^{-3 }=0.52\pm 0.08 \rm~ GeV cm^{-3}$  
that is compared with the other observations in Figure (\ref{fig-2}). We can calculate the local density of the dark matter from equation (\ref{eqnfw}) which is $\bar{\rho}_{0}=0.063 \pm 0.002 ~M_{\odot}\rm pc^{-3 }$, consistent with \cite{Widmark:2018ylf}.
\begin{table}
	\centering
	\caption{The local density of baryonic components at the Galactic midplane $(R_0,0)$.}					
	\begin{tabular}{|c | c| c| }
		\hline \hline
		$\rm Model$&$\rho_{g,0}(M_{\odot}\rm pc^{-3})$	&$\rho_{*,0}(M_{\odot}\rm pc^{-3})$\\
		\hline
		$\rm DM$&$ 0.049\pm 0.065$& $0.053\pm0.031$\\
		\hline
		$\rm MOG$&$0.050\pm0.046$& $0.055\pm0.040$\\
		\hline \hline
	\end{tabular}\label{tabrho}
\end{table}
\begin{figure}
	\begin{center}
		\includegraphics[width=7.5cm ]{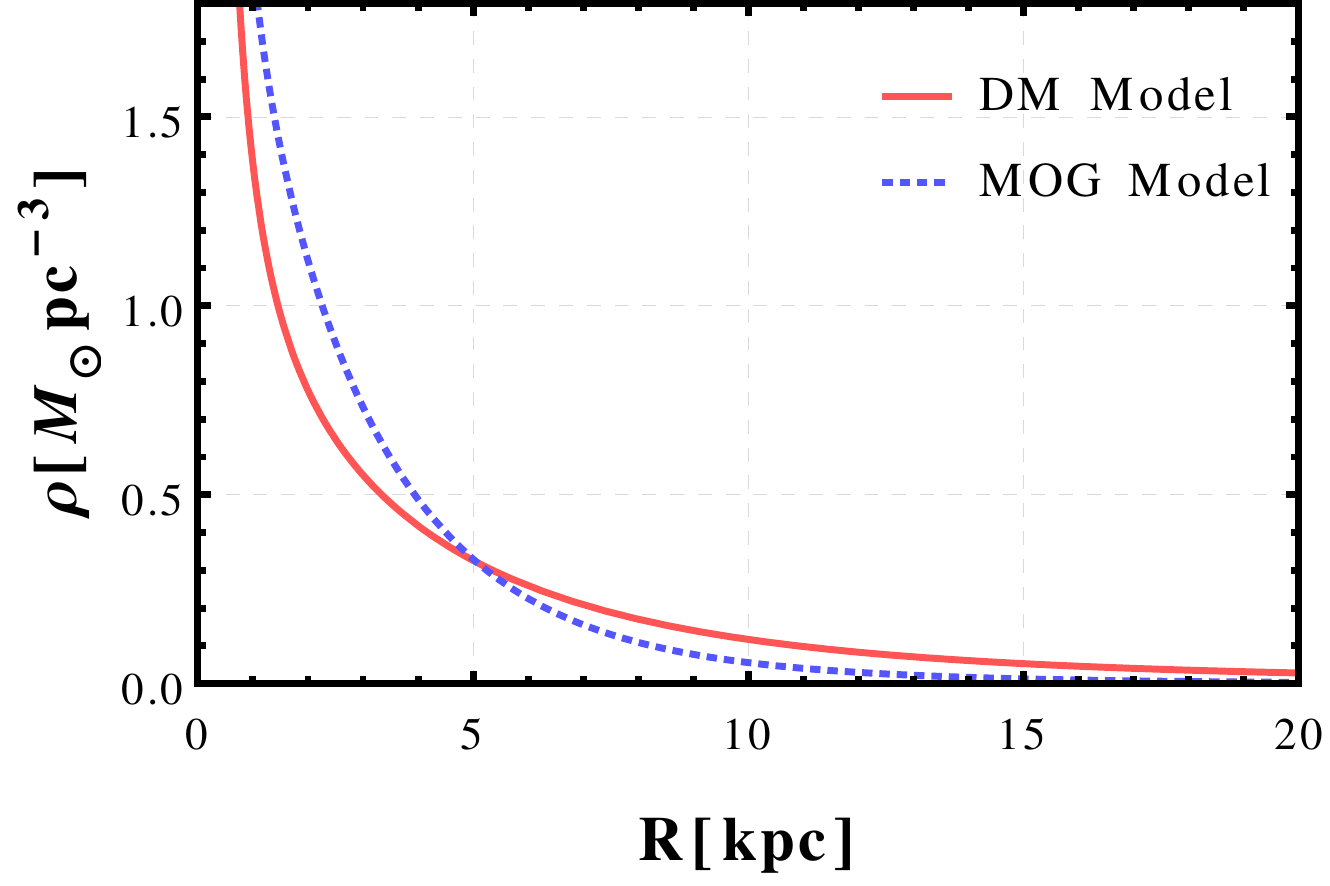}	
		\includegraphics[width=7.5cm ]{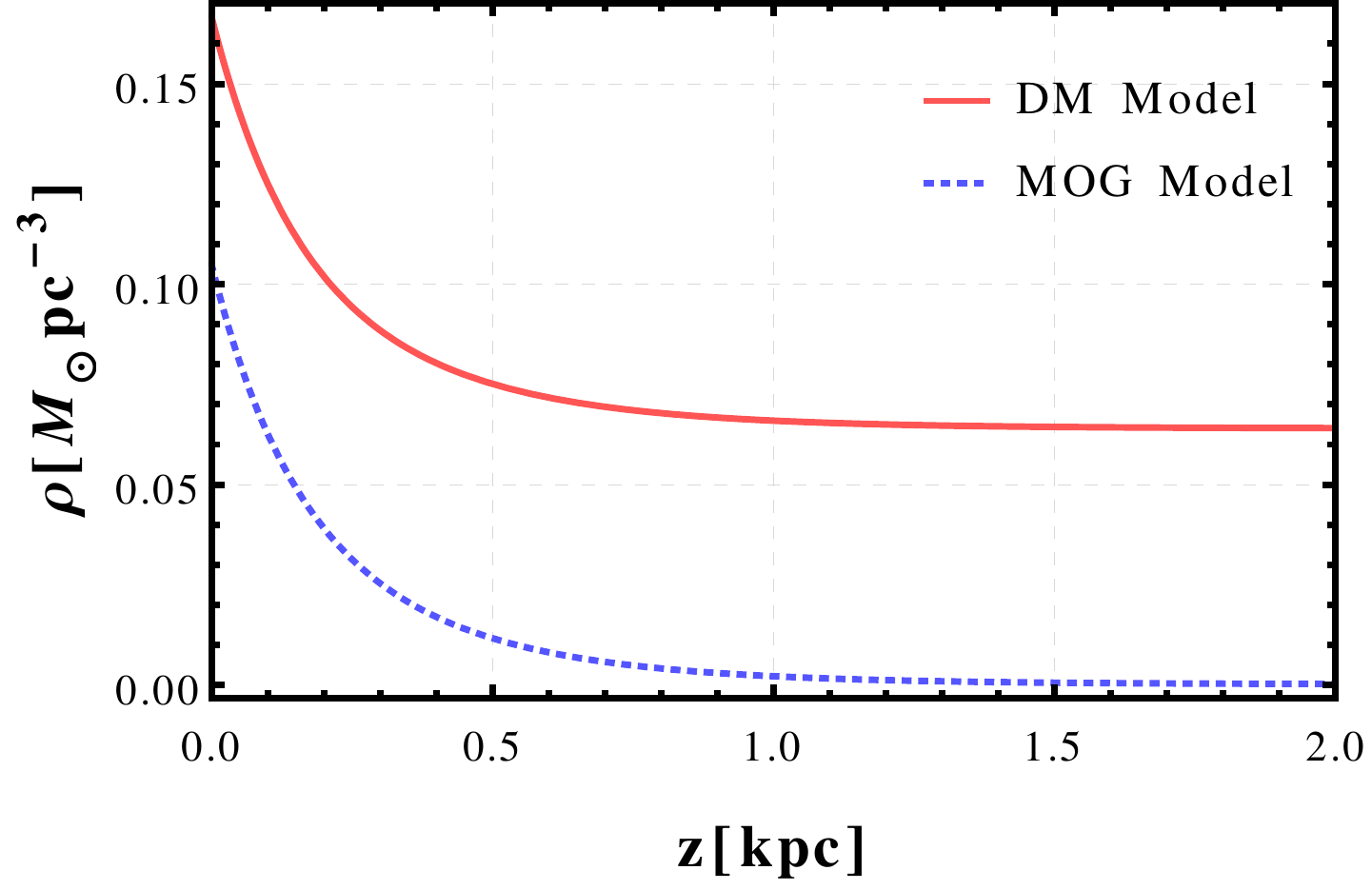}
		\caption{Total density profile of the MW as a function  of Galactocentric radius (upper panel) and the height above the Galactic plane (lower panel) at $R=8.12$ kpc.The solid red line represents the total matter contribution for DM model (i.e.  the sum of the bulge and the two disks as the baryonic counterpart plus the dark matter halo) and the blue dotted line represents the baryonic-matter for MOG model.}\label{fig-3}
	\end{center}
\end{figure}
\begin{figure}
	\begin{center}
		\includegraphics[width=8cm ]{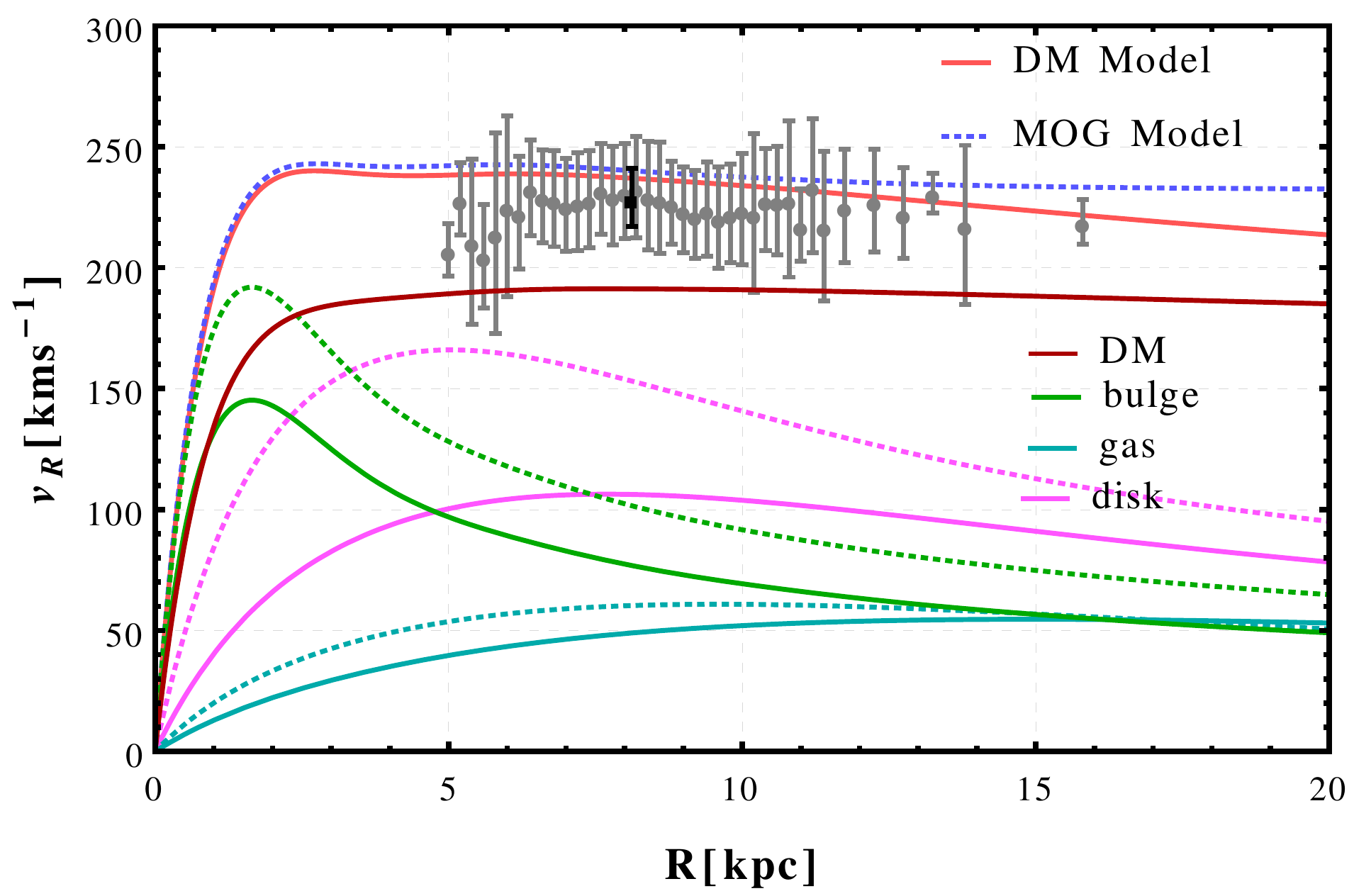}
		\caption{Rotation curves for DM and MOG models according to values given in Table (\ref{tabbest}). The black point correspond to $v_c (R_{0} )= 229.0 \pm 12 \rm km s^{-1}$ that is  determined in \citep{Eilers_2019}.The soiled curves for the DM model and the dashed curves for the MOG model. The contributions from bulge (green curves), disk (pink curves) and gas (cyan curves) are shown. }\label{fig-4}
	\end{center}
\end{figure}
Also, we let the parameters of the baryonic matter of the galaxy in equation (\ref{1}) and (\ref{rhodisk}) as the free parameters and obtain the local density of gas and stars of the disk. Table (\ref{tabrho}) presents the results of the baryonic densities and the local density of interstellar gas at the Galactic mid-plane for the two models. Our results in both models are consistent with the direct observation, $\rho_{g,0}=0.041\pm 0.004\rm M_{\odot}\rm pc^{-3 }$  in \cite{McKee:2015hwa}. For the two models, we calculate the total local baryonic density as 
 \begin{figure*}
 	\begin{center}
 		\includegraphics[width=5.5cm ]{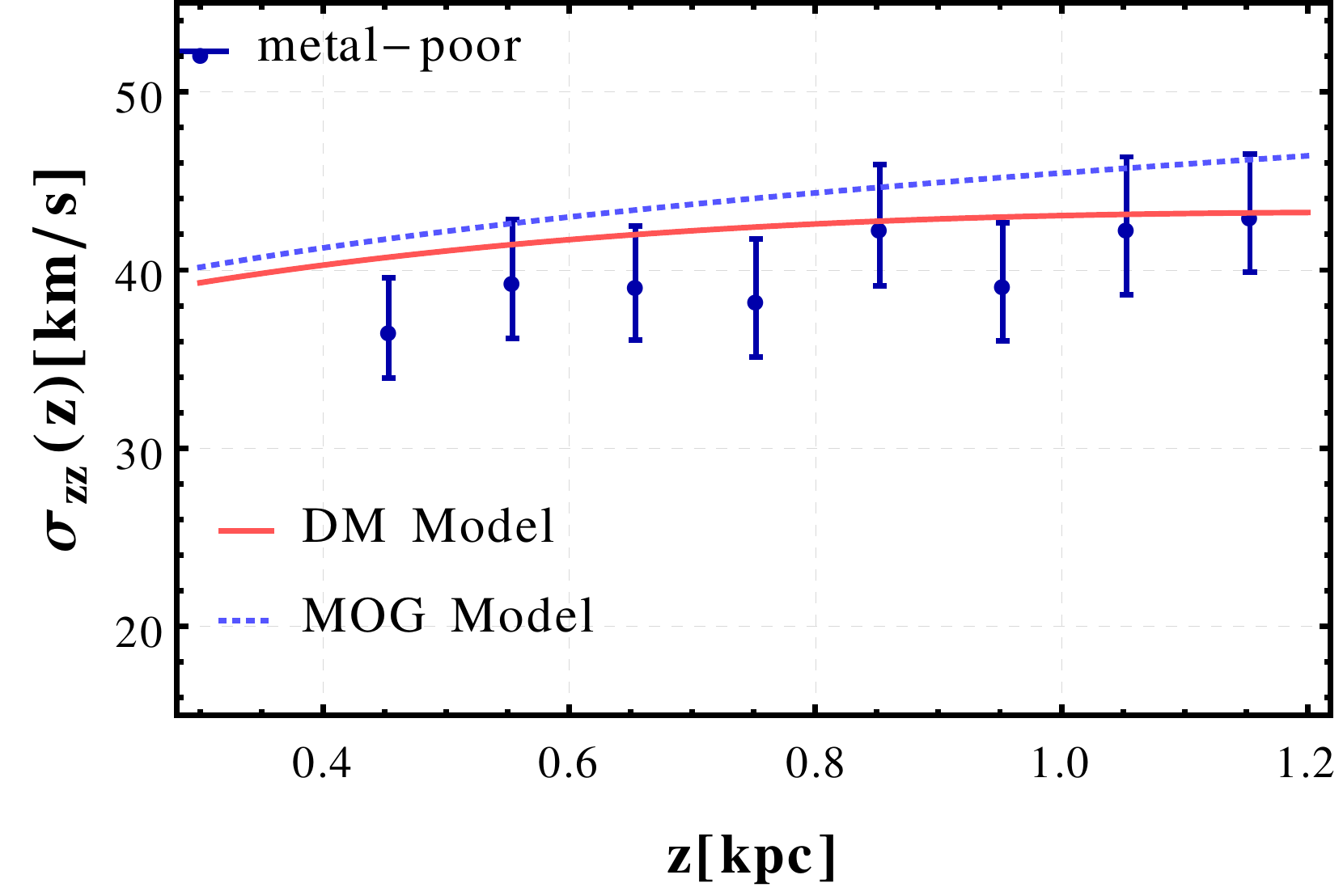}
 		\includegraphics[width=5.5cm ]{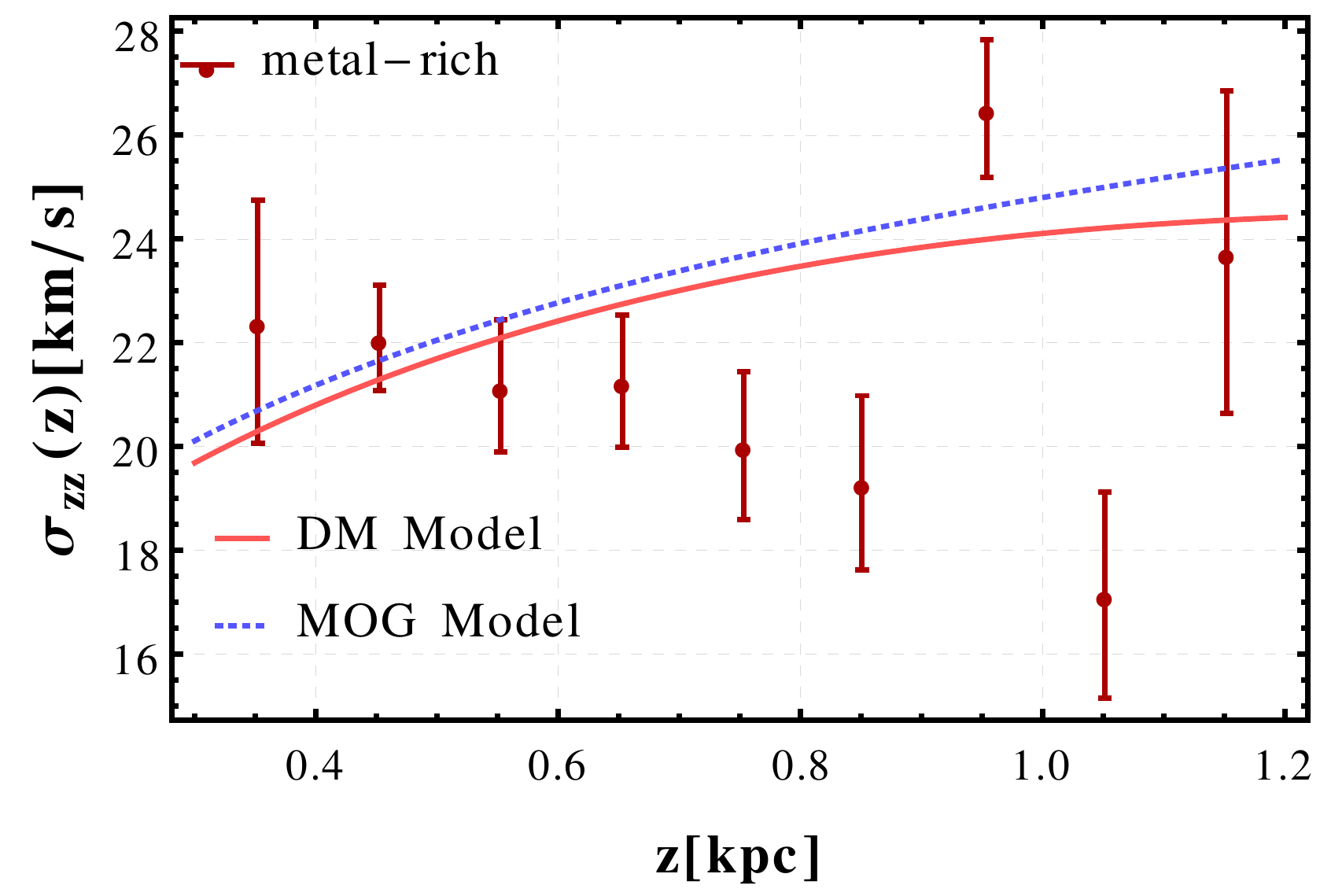}
 		\includegraphics[width=5.5cm ]{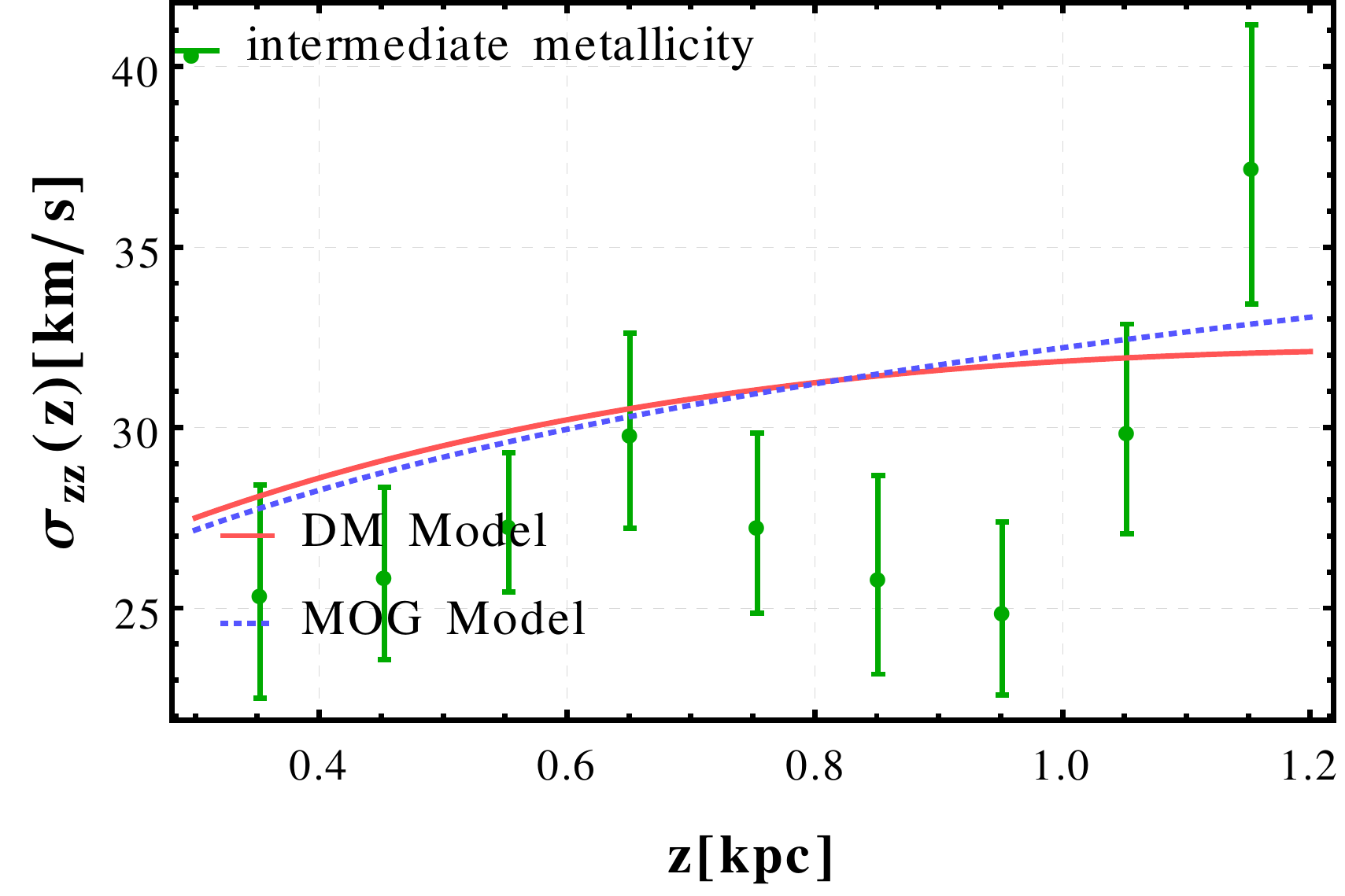}
 		\includegraphics[width=5.5cm ]{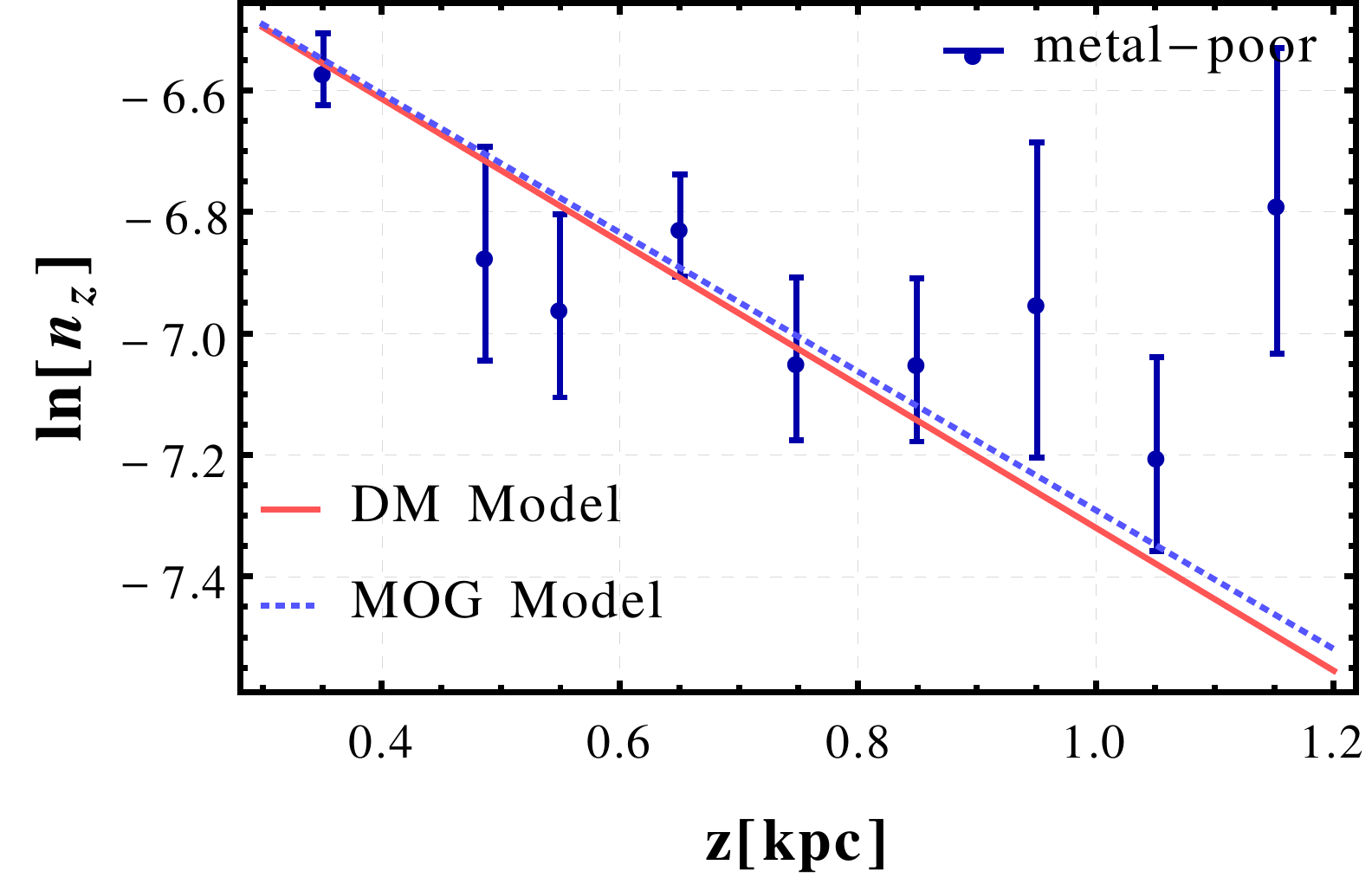}
 		\includegraphics[width=5.5cm ]{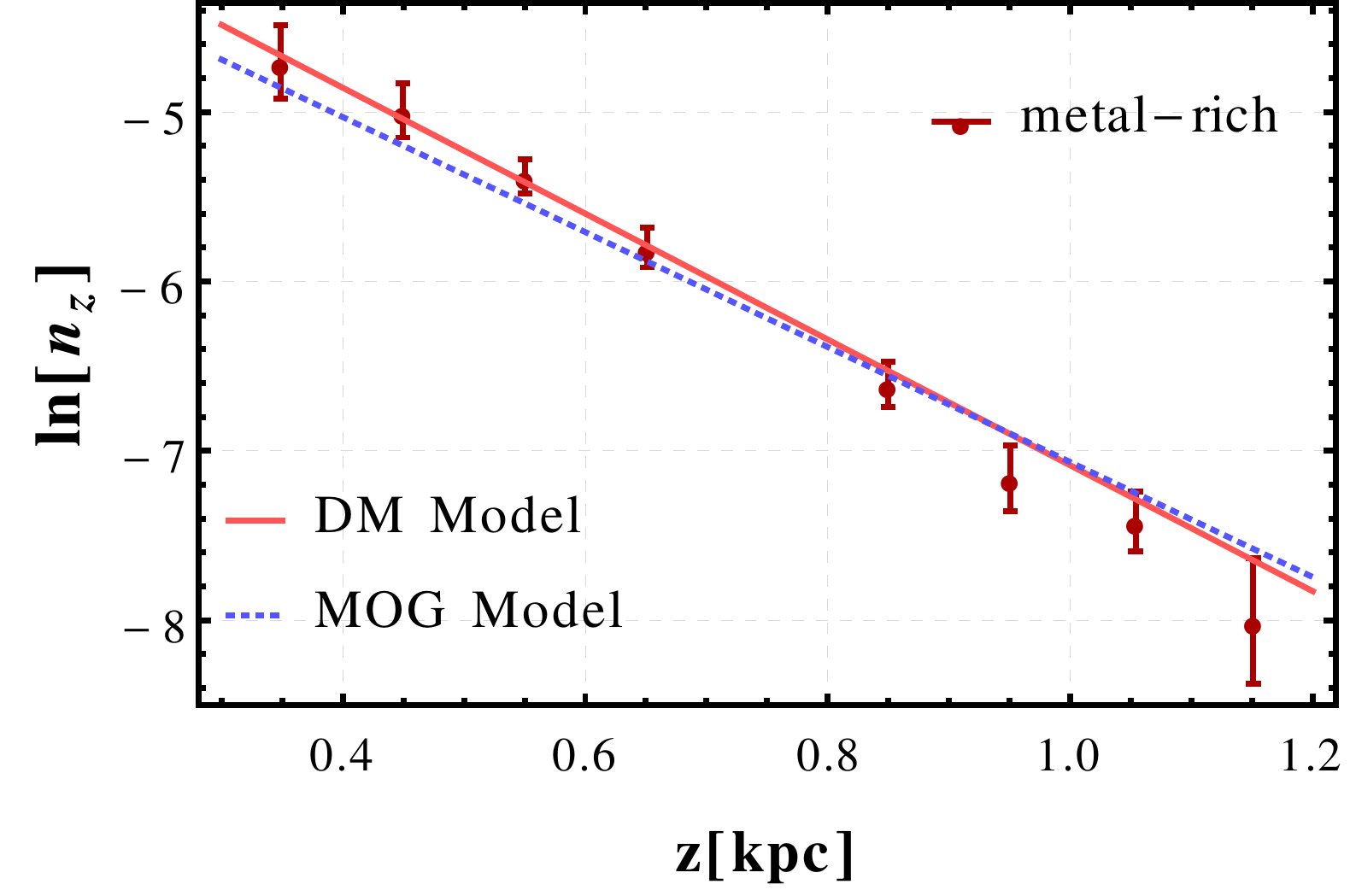}
 		\includegraphics[width=5.5cm ]{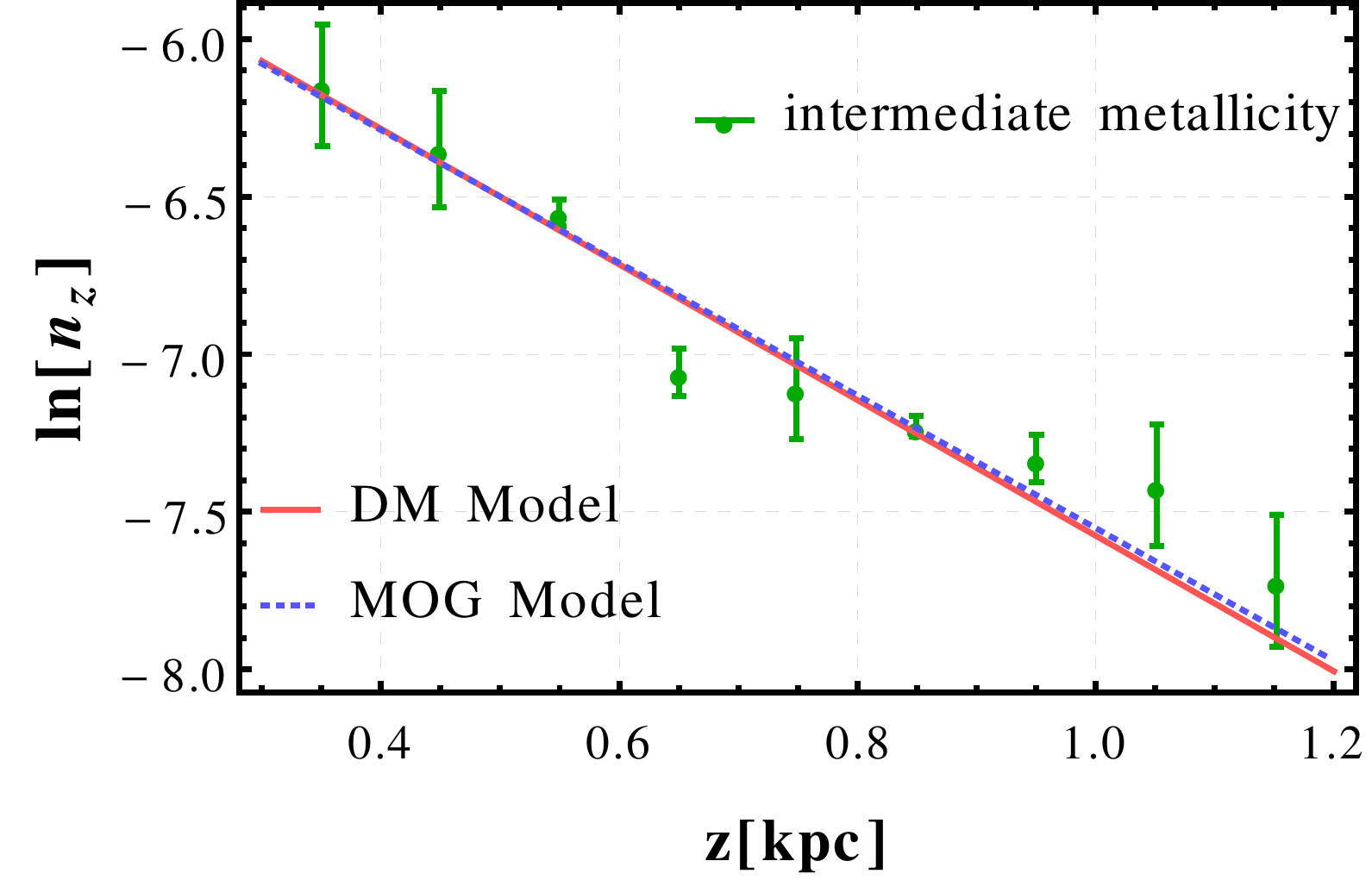}
 		\caption{Vertical velocity dispersions (top row) and number densities   (bottom row) according to values given in the of Table (\ref{tabbest}) for the three different stellar tracer populations used in this study.}\label{fig-5}
 	\end{center}
 \end{figure*}
 \begin{eqnarray}
 &&\rho_{\rm bar0}(DM)=0.103\pm 0.097 \rm M_{\odot}\rm pc^{-3 }\\
 &&\rho_{\rm bar0}(MOG)=0.105\pm 0.087 \rm M_{\odot}\rm pc^{-3 }.
 \end{eqnarray}
 The value of $ \rho_{\rm bar0}$ for two models we obtained  is also consistent with  \cite{Garbari:2012ff} with the value of $\rho_{\rm bar0}=0.098\pm 0.014 \rm M_{\odot}\rm pc^{-3 }$ and \cite{McKee:2015hwa}, with the value of  $\rho_{\rm bar0}=0.084\pm 0.012 \rm M_{\odot}\rm pc^{-3 }$ and \cite{Widmark:2018ylf} with the value of  $\rho_{\rm bar0}=0.0889\pm 0.0071 \rm M_{\odot}\rm pc^{-3 }$. 
 
 Combining the local dark matter and baryonic matter, we obtained the total density of matter as  $\rho_{\rm tot}=0.167 \pm 0.118 M_{\odot}\rm pc^{-3 }$. From our analyses in MOG, we also obtain $\alpha=8.99\pm0.02$ and $\mu=0.054\pm0.005$ kpc$^{-1}$ which is in agreement with the "universal" values of $\alpha=8.89\pm0.34$ and $\mu=0.042\pm0.004$ kpc$^{-1}$ \citep{Moffat:2013sja}. 
 In Figure (\ref{fig-3}), we show the matter density profile for the two models in terms of Galactocentric radius (R) and height above the Galactic plane (z) using the best-fit values of their parameters in Table (\ref{tabbest}). Since the density of disk exponentially decreases by distance and the density of halo in DM model decreases as a power-law function, for $R > 4$ kpc the density for DM model is larger than that of MOG model.\\
 In Figure (\ref{fig-4}), we use values in Table (\ref{tabbest}) and calculate the rotation curve of Galaxy for different components and the overall rotation velocity of DM and MOG. Our theoretical curves in MOG and DM are consistent with the data.\\
 Finally, we plotted vertical velocity dispersions and number density for the three different stellar tracer populations using the values of Table (\ref{tabbest}) in Figure (\ref{fig-5}). The dispersion velocity of the metal-poor stars has good compatibility of the DM and MOG with the observations. This class of stars is very old and they interact with the mean-field of the galaxy. In another word, the old stars forgot the memory of gravitational kicks from the other stars. However, the metal-rich and metal intermediate stars are young and do not follow the dynamics from the mean field of the Galaxy. The density of stars as a function of distance from the Galactic plane is almost consistent with both MOG and DM theories. 
\section{Conclusions}\label{sec4}
In this study, we compared the consistency of DM and MOG theories with local Milky Way observables including the rotation curve, the vertical dispersion velocity, the baryonic surface density, and the stellar disk profile. We described the baryonic mass distribution using a stellar disk, a gas disk, and a stellar bulge. For the DM scenario, the halo distribution was assumed to follow a spherical gNFW profile and for studying the behavior of MOG on astrophysical scales we used the weak field approximation of this theory. Under this scheme, the scalar fields remained constant and the acceleration of a test particle could obtain from the gradient of the effective potential.\\
We have performed a Bayesian likelihood analysis to compare two models and let the parameters $\{\tilde{\rho}_*,\tilde{\rho}_g,h_{*,R},M_{*,b},\tilde{\rho}_{dm},\gamma,R_s\}$ and $\{\tilde{\rho}_*,\tilde{\rho}_g,h_{*,R},M_{*,b},\alpha,\mu\}$ for DM and MOG models, respectively as the free parameters.
In comparison to MOG theory, DM model typically prefers smaller bulge mass, $M_{*,b}$ that consistent to photometric observations, $M_{*,b,obs}=(0.62\pm 0.31)\times10^{10}\rm M_{\odot}$ \citep{LopezCorredoira:2006na} but in tension with microlensing observations, $M_{*,b,obs}=(1.50\pm0.38)\times10^{10}\rm M_{\odot}$ \citep{Novatin:2007dd}. The best value from MOG is  $M_{*,b,MOG}=(1.06 \pm 0.26)\times10^{10}\rm M_{\odot}$ which is consistent with the microlensing result. 

We also obtained the local density of dark matter $\bar{\rho}_{dm}(R_0) = 0.0137\pm 0.002 \rm M_{\odot}\rm pc^{-3 }$ that is consistent with  $\bar{\rho}_{dm}(R_0) =0.0134 \pm 0.002 \rm M_{\odot}\rm pc^{-3 }$ by  \cite{Lin:2019yux},but it is a little bit larger than the $0.008 \pm 0.001 \rm M_{\odot}\rm pc^{-3 }$ value given by \cite{10.1093/mnras/stw2096}.
Two free parameters of MOG theory were obtained in this analysis are as
$\alpha=8.99\pm0.02$ and $\mu=0.054\pm0.005$ kpc$^{-1}$ which are in agreement with results of \cite{Moffat:2013sja}.

\section{Acknowledgments}
The work of ZD has been supported financially by  Iran Science Elites Federation.

		\bibliographystyle{mnras}
		\bibliography{ref}

\begin{thebibliography}{}
\makeatletter
\relax
\def\mn@urlcharsother{\let\do\@makeother \do\$\do\&\do\#\do\^\do\_\do\%\do\~}
\def\mn@doi{\begingroup\mn@urlcharsother \@ifnextchar [ {\mn@doi@}
  {\mn@doi@[]}}
\def\mn@doi@[#1]#2{\def\@tempa{#1}\ifx\@tempa\@empty \href
  {http://dx.doi.org/#2} {doi:#2}\else \href {http://dx.doi.org/#2} {#1}\fi
  \endgroup}
\def\mn@eprint#1#2{\mn@eprint@#1:#2::\@nil}
\def\mn@eprint@arXiv#1{\href {http://arxiv.org/abs/#1} {{\tt arXiv:#1}}}
\def\mn@eprint@dblp#1{\href {http://dblp.uni-trier.de/rec/bibtex/#1.xml}
  {dblp:#1}}
\def\mn@eprint@#1:#2:#3:#4\@nil{\def\@tempa {#1}\def\@tempb {#2}\def\@tempc
  {#3}\ifx \@tempc \@empty \let \@tempc \@tempb \let \@tempb \@tempa \fi \ifx
  \@tempb \@empty \def\@tempb {arXiv}\fi \@ifundefined
  {mn@eprint@\@tempb}{\@tempb:\@tempc}{\expandafter \expandafter \csname
  mn@eprint@\@tempb\endcsname \expandafter{\@tempc}}}

\bibitem[\protect\citeauthoryear{Abuter et~al.}{Abuter
  et~al.}{2018}]{Abuter:2018drb}
Abuter R.,  et~al., 2018, \mn@doi [Astron. Astrophys.]
  {10.1051/0004-6361/201833718}, 615, L15

\bibitem[\protect\citeauthoryear{Aghanim et~al.}{Aghanim
  et~al.}{2018}]{Aghanim:2018eyx}
Aghanim N.,  et~al., 2018

\bibitem[\protect\citeauthoryear{Aprile et~al.}{Aprile
  et~al.}{2018}]{Aprile:2018dbl}
Aprile E.,  et~al., 2018, \mn@doi [Phys. Rev. Lett.]
  {10.1103/PhysRevLett.121.111302}, 121, 111302

\bibitem[\protect\citeauthoryear{Atwood et~al.}{Atwood
  et~al.}{2009}]{Atwood:2009ez}
Atwood W.~B.,  et~al., 2009, \mn@doi [Astrophys. J.]
  {10.1088/0004-637X/697/2/1071}, 697, 1071

\bibitem[\protect\citeauthoryear{Bekenstein}{Bekenstein}{2004}]{Bekenstein:2004ne}
Bekenstein J.~D.,  2004, \mn@doi [Phys. Rev.] {10.1103/PhysRevD.70.083509,
  10.1103/PhysRevD.71.069901}, D70, 083509

\bibitem[\protect\citeauthoryear{Benito}{Benito}{2019}]{Benito:2019vef}
Benito M.,  2019, PhD thesis, Sao Paulo, IFT

\bibitem[\protect\citeauthoryear{Bienaymé et~al.}{Bienaymé
  et~al.}{2014}]{Bienayme:2014kva}
Bienaymé O.,  et~al., 2014, \mn@doi [Astron. Astrophys.]
  {10.1051/0004-6361/201424478}, 571, A92

\bibitem[\protect\citeauthoryear{Bland~Hawthorn \& Gerhard}{Bland~Hawthorn \&
  Gerhard}{2016}]{Bland:2017}
Bland~Hawthorn J.,  Gerhard O.,  2016, \mn@doi [Annual Reviews of Astronomy and
  Astrophysics.] {10.1146/annurev-astro-081915-023441}, 54, 529

\bibitem[\protect\citeauthoryear{Bonilla~Rivera \&
  García-Farieta}{Bonilla~Rivera \& García-Farieta}{2019}]{Rivera:2016zzr}
Bonilla~Rivera A.,  García-Farieta J.~E.,  2019, \mn@doi [Int. J. Mod. Phys.]
  {10.1142/S0218271819501189}, D28, 1950118

\bibitem[\protect\citeauthoryear{Bovy}{Bovy}{2017}]{Bovy}
Bovy J.,  2017, \mn@doi [Mon. Not. Roy. Astron. Soc.] {10.1093/mnrasl/slx027},
  468, 63

\bibitem[\protect\citeauthoryear{Calchi~Novati, De~Luca, Jetzer, Mancini  \&
  Scarpetta}{Calchi~Novati et~al.}{2008}]{Novatin:2007dd}
Calchi~Novati S.,  De~Luca F.,  Jetzer P.,  Mancini L.,   Scarpetta G.,  2008,
  \mn@doi [Astron. Astrophys.] {10.1051/0004-6361:20078439}, 480, 723

\bibitem[\protect\citeauthoryear{Crosta, Giammaria, Lattanzi  \& Poggio}{Crosta
  et~al.}{2018}]{Crosta:2018var}
Crosta M.,  Giammaria M.,  Lattanzi M.~G.,   Poggio E.,  2018

\bibitem[\protect\citeauthoryear{Dutta \& Islam}{Dutta \&
  Islam}{2018}]{Dutta:2018oaj}
Dutta K.,  Islam T.,  2018, \mn@doi [Phys. Rev. D]
  {10.1103/PhysRevD.98.124012}, 98, 124012

\bibitem[\protect\citeauthoryear{Eilers, Hogg, Rix  \& Ness}{Eilers
  et~al.}{2019}]{Eilers_2019}
Eilers A.-C.,  Hogg D.~W.,  Rix H.-W.,   Ness M.~K.,  2019, \mn@doi [The
  Astrophysical Journal] {10.3847/1538-4357/aaf648}, 871, 120

\bibitem[\protect\citeauthoryear{Einasto}{Einasto}{1965}]{Einasto:1965czb}
Einasto J.,  1965, Trudy Astrofizicheskogo Instituta Alma-Ata, 5, 87

\bibitem[\protect\citeauthoryear{Garbari, Liu, Read  \& Lake}{Garbari
  et~al.}{2012}]{Garbari:2012ff}
Garbari S.,  Liu C.,  Read J.~I.,   Lake G.,  2012, \mn@doi [Mon. Not. Roy.
  Astron. Soc.] {10.1111/j.1365-2966.2012.21608.x}, 425, 1445

\bibitem[\protect\citeauthoryear{Hernquist}{Hernquist}{1990}]{Hernquist:1990be}
Hernquist L.,  1990, \mn@doi [Astrophys. J.] {10.1086/168845}, 356, 359

\bibitem[\protect\citeauthoryear{Huang et~al.,}{Huang
  et~al.}{2016}]{10.1093/mnras/stw2096}
Huang Y.,  et~al., 2016, \mn@doi [Monthly Notices of the Royal Astronomical
  Society] {10.1093/mnras/stw2096}, 463, 2623

\bibitem[\protect\citeauthoryear{Iocco, Pato  \& Bertone}{Iocco
  et~al.}{2015a}]{Iocco:2015xga}
Iocco F.,  Pato M.,   Bertone G.,  2015a, \mn@doi [Nature Phys.]
  {10.1038/nphys3237}, 11, 245

\bibitem[\protect\citeauthoryear{Iocco, Pato  \& Bertone}{Iocco
  et~al.}{2015b}]{Iocco:2015iia}
Iocco F.,  Pato M.,   Bertone G.,  2015b, \mn@doi [Phys. Rev.]
  {10.1103/PhysRevD.92.084046}, D92, 084046

\bibitem[\protect\citeauthoryear{Islam \& Dutta}{Islam \&
  Dutta}{2020}]{Islam:2019iua}
Islam T.,  Dutta K.,  2020, \mn@doi [Phys. Rev. D]
  {10.1103/PhysRevD.101.084015}, 101, 084015

\bibitem[\protect\citeauthoryear{J.~Binney}{J.~Binney}{1998}]{Binney1998}
J.~Binney M.~M.,  1998, Galactic astronomy.
Princeton University Press

\bibitem[\protect\citeauthoryear{J.~Binney}{J.~Binney}{2008}]{Binney2008}
J.~Binney S.~T.,  2008, Galactic Dynamics.
Princeton University Press

\bibitem[\protect\citeauthoryear{Karukes, Benito, Iocco, Trotta  \&
  Geringer-Sameth}{Karukes et~al.}{2019}]{Karukes:2019jxv}
Karukes E.~V.,  Benito M.,  Iocco F.,  Trotta R.,   Geringer-Sameth A.,  2019,
  \mn@doi [JCAP] {10.1088/1475-7516/2019/09/046}, 1909, 046

\bibitem[\protect\citeauthoryear{Klypin, Kravtsov, Valenzuela  \& Prada}{Klypin
  et~al.}{1999}]{Klypin:1999uc}
Klypin A.~A.,  Kravtsov A.~V.,  Valenzuela O.,   Prada F.,  1999, \mn@doi
  [Astrophys. J.] {10.1086/307643}, 522, 82

\bibitem[\protect\citeauthoryear{Kuijken \& Gilmore}{Kuijken \&
  Gilmore}{1989}]{Kuijken:1989hu}
Kuijken K.,  Gilmore G.,  1989, Mon. Not. Roy. Astron. Soc., 239, 605

\bibitem[\protect\citeauthoryear{Lin \& Li}{Lin \& Li}{2019}]{Lin:2019yux}
Lin H.-N.,  Li X.,  2019, \mn@doi [Mon. Not. Roy. Astron. Soc.]
  {10.1093/mnras/stz1698}, 487, 5679

\bibitem[\protect\citeauthoryear{Lisanti, Moschella, Outmezguine  \&
  Slone}{Lisanti et~al.}{2018}]{Lisanti:2018qam}
Lisanti M.,  Moschella M.,  Outmezguine N.~J.,   Slone O.,  2018

\bibitem[\protect\citeauthoryear{Lopez-Corredoira, Cabrera-Lavers, Mahoney,
  Hammersley, Garzon  \& Gonzalez-Fernandez}{Lopez-Corredoira
  et~al.}{2007}]{LopezCorredoira:2006na}
Lopez-Corredoira M.,  Cabrera-Lavers A.,  Mahoney T.~J.,  Hammersley P.~L.,
  Garzon F.,   Gonzalez-Fernandez C.,  2007, \mn@doi [Astron. J.]
  {10.1086/509605}, 133, 154

\bibitem[\protect\citeauthoryear{Maleki, Baghram  \& Rahvar}{Maleki
  et~al.}{2019}]{Maleki:2019xya}
Maleki A.,  Baghram S.,   Rahvar S.,  2019

\bibitem[\protect\citeauthoryear{McKee, Parravano  \& Hollenbach}{McKee
  et~al.}{2015}]{McKee:2015hwa}
McKee C.~F.,  Parravano A.,   Hollenbach D.~J.,  2015, \mn@doi [Astrophys. J.]
  {10.1088/0004-637X, 10.1088/0004-637X/814/1/13}, 814, 13

\bibitem[\protect\citeauthoryear{McMillan}{McMillan}{2011}]{McMillan:2011wd}
McMillan P.~J.,  2011, \mn@doi [Mon. Not. Roy. Astron. Soc.]
  {10.1111/j.1365-2966.2011.18564.x}, 414, 2446

\bibitem[\protect\citeauthoryear{Milgrom}{Milgrom}{1983}]{Milgrom:1983ca}
Milgrom M.,  1983, \mn@doi [Astrophys. J.] {10.1086/161130}, 270, 365

\bibitem[\protect\citeauthoryear{Misiriotis, Xilouris, Papamastorakis, Boumis
  \& Goudis}{Misiriotis et~al.}{2006}]{Misiriotis:2006qq}
Misiriotis A.,  Xilouris E.~M.,  Papamastorakis J.,  Boumis P.,   Goudis C.~D.,
   2006, \mn@doi [Astron. Astrophys.] {10.1051/0004-6361:20054618}, 459, 113

\bibitem[\protect\citeauthoryear{Widmark}{Mof}{}]{Moffat:2013sja}


\bibitem[\protect\citeauthoryear{Moffat}{Moffat}{2006}]{Moffat:2005si}
Moffat J.~W.,  2006, \mn@doi [JCAP] {10.1088/1475-7516/2006/03/004}, 0603, 004

\bibitem[\protect\citeauthoryear{{Moffat} \& {Rahvar}}{{Moffat} \&
  {Rahvar}}{2013}]{rahvar1}
{Moffat} J.~W.,  {Rahvar} S.,  2013, \mn@doi [\mnras] {10.1093/mnras/stt1670},
  \href {https://ui.adsabs.harvard.edu/abs/2013MNRAS.436.1439M} {436, 1439}

\bibitem[\protect\citeauthoryear{{Moffat} \& {Rahvar}}{{Moffat} \&
  {Rahvar}}{2014}]{rahvar2}
{Moffat} J.~W.,  {Rahvar} S.,  2014, \mn@doi [\mnras] {10.1093/mnras/stu855},
  \href {https://ui.adsabs.harvard.edu/abs/2014MNRAS.441.3724M} {441, 3724}

\bibitem[\protect\citeauthoryear{Moffat \& Toth}{Moffat \&
  Toth}{2009}]{Moffat:2007nj}
Moffat J.,  Toth V.,  2009, \mn@doi [Class. Quant. Grav.]
  {10.1088/0264-9381/26/8/085002}, 26, 085002

\bibitem[\protect\citeauthoryear{Moffat \& Toth}{Moffat \&
  Toth}{2015}]{Moffat:2014pia}
Moffat J.,  Toth V.,  2015, \mn@doi [Phys. Rev. D]
  {10.1103/PhysRevD.91.043004}, 91, 043004

\bibitem[\protect\citeauthoryear{Moore, Quinn, Governato, Stadel  \&
  Lake}{Moore et~al.}{1999}]{Moore:1999gc}
Moore B.,  Quinn T.~R.,  Governato F.,  Stadel J.,   Lake G.,  1999, \mn@doi
  [Mon. Not. Roy. Astron. Soc.] {10.1046/j.1365-8711.1999.03039.x}, 310, 1147

\bibitem[\protect\citeauthoryear{Navarro, Frenk  \& White}{Navarro
  et~al.}{1996}]{Navarro:1995iw}
Navarro J.~F.,  Frenk C.~S.,   White S. D.~M.,  1996, \mn@doi [Astrophys. J.]
  {10.1086/177173}, 462, 563

\bibitem[\protect\citeauthoryear{Navarro et~al.,}{Navarro
  et~al.}{2004}]{Navarro:2003ew}
Navarro J.~F.,  et~al., 2004, \mn@doi [Mon. Not. Roy. Astron. Soc.]
  {10.1111/j.1365-2966.2004.07586.x}, 349, 1039

\bibitem[\protect\citeauthoryear{Negrelli, Benito, Landau, Iocco  \&
  Kraiselburd}{Negrelli et~al.}{2018}]{Negrelli:2018gll}
Negrelli C.,  Benito M.,  Landau S.,  Iocco F.,   Kraiselburd L.,  2018,
  \mn@doi [Phys. Rev. D] {10.1103/PhysRevD.98.104061}, 98, 104061

\bibitem[\protect\citeauthoryear{Nesti \& Salucci}{Nesti \&
  Salucci}{2013}]{Nesti:2013uwa}
Nesti F.,  Salucci P.,  2013, \mn@doi [JCAP] {10.1088/1475-7516/2013/07/016},
  1307, 016

\bibitem[\protect\citeauthoryear{Nieuwenhuizen}{Nieuwenhuizen}{2017}]{Nieuwenhuizen:2016uxv}
Nieuwenhuizen T.~M.,  2017, \mn@doi [Fortsch. Phys.] {10.1002/prop.201600050},
  65, 1600050

\bibitem[\protect\citeauthoryear{Pato \& Iocco}{Pato \&
  Iocco}{2017}]{Pato:2017yai}
Pato M.,  Iocco F.,  2017, ] {10.1016/j.softx.2016.12.006}

\bibitem[\protect\citeauthoryear{Pato, Iocco  \& Bertone}{Pato
  et~al.}{2015}]{Pato:2015dua}
Pato M.,  Iocco F.,   Bertone G.,  2015, \mn@doi [JCAP]
  {10.1088/1475-7516/2015/12/001}, 1512, 001

\bibitem[\protect\citeauthoryear{Rebassa-Mansergas et~al.,}{Rebassa-Mansergas
  et~al.}{2016}]{Rebassa_Mansergas_2016}
Rebassa-Mansergas A.,  et~al., 2016, \mn@doi [Monthly Notices of the Royal
  Astronomical Society] {10.1093/mnras/stw2021}, 463, 1137–1143

\bibitem[\protect\citeauthoryear{Rubin \& Ford}{Rubin \&
  Ford}{1970}]{Rubin:1970zza}
Rubin V.~C.,  Ford Jr. W.~K.,  1970, \mn@doi [Astrophys. J.] {10.1086/150317},
  159, 379

\bibitem[\protect\citeauthoryear{Rubin, Thonnard  \& Ford}{Rubin
  et~al.}{1980}]{Rubin:1980zd}
Rubin V.~C.,  Thonnard N.,   Ford Jr. W.~K.,  1980, \mn@doi [Astrophys. J.]
  {10.1086/158003}, 238, 471

\bibitem[\protect\citeauthoryear{Salucci, Nesti, Gentile  \& Martins}{Salucci
  et~al.}{2010}]{Salucci:2010qr}
Salucci P.,  Nesti F.,  Gentile G.,   Martins C.~F.,  2010, \mn@doi [Astron.
  Astrophys.] {10.1051/0004-6361/201014385}, 523, A83

\bibitem[\protect\citeauthoryear{Schutz, Lin, Safdi  \& Wu}{Schutz
  et~al.}{2018}]{Schutz:2017tfp}
Schutz K.,  Lin T.,  Safdi B.~R.,   Wu C.-L.,  2018, \mn@doi [Phys. Rev. Lett.]
  {10.1103/PhysRevLett.121.081101}, 121, 081101

\bibitem[\protect\citeauthoryear{Scolnic et~al.}{Scolnic
  et~al.}{2017}]{Scolnic:2017caz}
Scolnic D.~M.,  et~al., 2017, ] {10.17909/T95Q4X}

\bibitem[\protect\citeauthoryear{Sofue}{Sofue}{2012}]{10.1093/pasj/64.4.75}
Sofue Y.,  2012, \mn@doi [Publications of the Astronomical Society of Japan]
  {10.1093/pasj/64.4.75}, 64

\bibitem[\protect\citeauthoryear{Widmark}{Widmark}{2019}]{Widmark:2018ylf}
Widmark A.,  2019, \mn@doi [Astron. Astrophys.] {10.1051/0004-6361/201834718},
  623, A30

\bibitem[\protect\citeauthoryear{Wielen}{Wielen}{1977}]{Wielen:1977zz}
Wielen R.,  1977, Astron. Astrophys., 60, 263

\bibitem[\protect\citeauthoryear{Zhang, Rix, van~de Ven, Bovy, Liu  \&
  Zhao}{Zhang et~al.}{2013}]{Zhang:2012rsb}
Zhang L.,  Rix H.-W.,  van~de Ven G.,  Bovy J.,  Liu C.,   Zhao G.,  2013,
  \mn@doi [Astrophys. J.] {10.1088/0004-637X/772/2/108}, 772, 108

\bibitem[\protect\citeauthoryear{Zhao et~al.}{Zhao et~al.}{2019}]{Zhao:2018jxv}
Zhao G.-B.,  et~al., 2019, \mn@doi [Mon. Not. Roy. Astron. Soc.]
  {10.1093/mnras/sty2845}, 482, 3497

\bibitem[\protect\citeauthoryear{Zwicky}{Zwicky}{1937}]{Zwicky:1937zza}
Zwicky F.,  1937, \mn@doi [Astrophys. J.] {10.1086/143864}, 86, 217

\makeatother
\end{thebibliography}
		
	\end{document}